\documentclass{aa}  

\usepackage{graphicx}
\usepackage{longtable}
\usepackage{txfonts}
\begin{document}

   \title{Testing the accretion scenario of $\lambda$ Boo stars}

   
   \titlerunning{Testing the accretion scenario of $\lambda$ Boo stars}
   \authorrunning{Saffe et al.}

   \author{J. Alacoria\inst{1,6}, C. Saffe\inst{1,2,6}, M. Jaque Arancibia\inst{3,4}, R. Angeloni\inst{5}, 
           P. Miquelarena\inst{1,2,6}, M. Flores\inst{1,2,6}, M. E. Veramendi\inst{1,6,7},
            \and A. Collado\inst{1,2,6}
           }

\institute{Instituto de Ciencias Astron\'omicas, de la Tierra y del Espacio (ICATE-CONICET), C.C 467, 5400, San Juan, Argentina.
         \and Universidad Nacional de San Juan (UNSJ), Facultad de Ciencias Exactas, F\'isicas y Naturales (FCEFN), San Juan, Argentina.
         \and{Instituto de Investigaci\'on Multidisciplinar en Ciencia y Tecnolog\'ia, Universidad de La Serena, Ra\'ul Bitr\'an 1305, La Serena, Chile}
         \and Departamento de F\'isica y Astronom\'ia, Universidad de La Serena, Av. Cisternas 1200 N, La Serena, Chile.         
         \and Gemini Observatory / NSF’s NOIRLab, Casilla 603, La Serena, Chile
        \and Consejo Nacional de Investigaciones Cient\'ificas y T\'ecnicas (CONICET), Argentina
        \and Universidad Nacional de San Juan (UNSJ), Facultad de Filosofía, Humanidades y Artes (FFHA), San Juan, Argentina
         }

   \date{Received xxx, xxx ; accepted xxxx, xxxx}

 
  \abstract
   {The group of $\lambda$ Boo stars is known for years, however the origin of its chemical peculiarity is still strongly debated.}
   {Our aim is to test the accretion scenario of $\lambda$ Boo stars.
   This model predicts that a binary system with two early-type stars passing through a diffuse cloud should both display the same
   superficial peculiarity. 
   }
   {Via spectral synthesis, we carried out a detailed abundance determination of three multiple systems hosting a candidate $\lambda$ Boo
    star: the remarkable triple system HD 15164/65/65C and the two binary systems HD 193256/281 and HD 198160/161.
   Stellar parameters were initially estimated using Str\"omgren photometry or literature values and then
   refined by requiring excitation and ionization balances of Fe lines.
   The abundances were determined iteratively for 24 different species by fitting synthetic spectra using the SYNTHE program
   together with local thermodynamic equilibrium (LTE) ATLAS12 model atmospheres.
   Specific opacities were calculated for each star, depending on its arbitrary composition and microturbulence velocity
   v$_\mathrm{micro}$ through the opacity sampling (OS) method. 
   The abundances of the light elements C and O were corrected by Non-LTE effects.
   The complete chemical pattern of the stars were then compared to those of $\lambda$ Boo stars.}
   {
   The abundance analysis of the triple system HD 15164/65/65C shows a clear $\lambda$ Boo object (HD 15165)
   and two objects with near solar composition (HD 15164 and 15165C).
   Notably, the presence of a $\lambda$ Boo star (HD 15165) together with a near solar early-type object (HD 15164) is
   difficult to explain under
   the accretion scenario. Also, the solar-like composition derived for the late-type star of the system (HD 15165C)
   could be used, for the first time, as a proxy for the initial composition of the $\lambda$ Boo stars. This could help to constrain
   any model of $\lambda$ Boo stars formation and not only the accretion scenario.
   The abundance analysis of the binary system HD 193256/281 shows no clear $\lambda$ Boo components, while the analysis of
   HD 198160/161 shows two mild-$\lambda$ Boo stars.
   Then, by carefully reviewing abundance analysis of all known binary systems with candidate $\lambda$ Boo stars from literature
   and including the systems analyzed here, we find no binary/multiple system having two clear "bonafide" $\lambda$ Boo stars,
   as expected from the accretion scenario.
   The closer candidates to show two $\lambda$ Boo-like stars are the binary systems HD 84948, HD 171948 and HD 198160;
   however, in our opinion they show mild rather than clear $\lambda$ Boo patterns.
   }
   {We performed for the first time a complete analysis of a triple system which includes a $\lambda$ Boo candidate.
   Our results brings little support to the accretion scenario of $\lambda$ Boo stars. Then,
   there is an urgent need of additional binary and multiple systems to be analyzed through a detailed abundance analysis,
   in order to test the accretion model of $\lambda$ Boo stars.
   }
   
   \keywords{Stars: abundances -- 
             Stars: binaries -- 
             Stars: chemically peculiar -- 
             Stars: individual: {HD 15164, HD 15165, HD 15165C, HD 193256, HD 193281, HD 198160, HD 198161}
            }

   \maketitle
%

\section{Introduction}

The main feature of $\lambda$ Boo stars is a notable underabundance of most Fe-peak elements
and near solar abundances of lighter elements (C, N, O and S).
They comprise main-sequence late-B to early-F stars, where a maximum of about 2\% of all objects
are believed to be $\lambda$ Boo stars \citep{gray-corbally98,paunzen01b}.
Classification-resolution spectroscopy shows promising $\lambda$ Boo candidates \citep[e.g.,][]{murphy15,gray17},
and a more detailed abundance determination, especially including the lighter elements,
is considered a ultimate test to confirm that a candidate is indeed a bonafide member of the class
\citep[e.g.,][]{andrievsky02,heiter02}.

The origin of the $\lambda$ Boo peculiarity still remains as a challenge 
\citep[see, e.g., the recent discussion of ][and references therein]{murphy-paunzen17}.
Their rotational velocities do not necessarily point toward lower values, 
marking thus a difference with chemically peculiar Am and Ap stars \citep{abt-morrell95,murphy15}.
A possible explanation consist in the interaction of the star with a diffuse interstellar cloud 
\citep{kamp-paunzen02,mg09}.
In this work, we refer to this model as the "accretion scenario", in which 
the underabundances are produced by different amounts of volatile accreted material,
and the more refractory species are possibly separated and repelled from the star.
More recently, \citet{jura15} proposed that this peculiar pattern possibly originates from the 
winds of hot-Jupiter planets\footnote{Hot-Jupiter planets present short orbital periods ($<$10 d) and large masses ($>$ 0.1 M$_{Jup}$).}.
In this case, the planet acts as the source of gas poor in refractory species. 
However, \citet{saffe21} have recently shown that eight early-type stars hosting hot-Jupiter planets
do not display the $\lambda$ Boo peculiarity.
This would let the interaction of the star with a diffuse cloud, as the more plausible
scenario to explain the $\lambda$ Boo phenomena in main-sequence stars.

Under the accretion scenario, two early-type stars passing through a diffuse cloud should display, in principle,
the same superficial peculiarity \citep[e.g.,][]{paunzen12a,paunzen12b}. At the same time, hotter stars 
(T$_\mathrm{eff}$ $>$ $\sim$12000 K) with strong winds, and cooler stars (T$_\mathrm{eff}$ $<$ $\sim$6500 K)
with larger convective zones, should not notably change their composition.
These predictions make the analysis of binary and multiple systems an important tool to test the accretion scenario.
However, the number of known candidate $\lambda$ Boo stars in binary/multiple systems is limited to a dozen of objects
\citep[e.g.,][]{paunzen12a,paunzen12b}, where most of them are spectroscopic binary (SB) systems.
To our knowledge, only five of these systems present a detailed chemical analysis of the two components 
(see the Appendix for a more detailed review).
Notably, some stars of these binary systems were recently identified as non-members or uncertain members of the $\lambda$ Boo class
\citep[see ][]{gray17}.
Based on literature data, we selected for this study three binary/multiple systems 
that possibly confront the accretion scenario. In addition, they are spatially resolved
\citep[in contrast to most candidate $\lambda$ Boo stars that belong to SB systems, ][]{paunzen12a,paunzen12b},
allowing a individual analysis without a strong contribution from the companion.
We also review all known binary or multiple systems with candidate $\lambda$ Boo stars,
with data taken from the literature (see Appendix).

In this work, we present an analysis of the remarkable triple system HD 15165.
It is composed by HD 15165, HD 15164 and HD 15165C (stars A, B and C) 
with spectral types "F2 V kA2mA2 $\lambda$ Boo?", "F1 V kA7mA6 ($\lambda$ Boo)?" and "K2 V" \citep{murphy15}.
Some previous works suggest that the A star belong to the $\lambda$ Boo class \citep{andrievsky95,cherny98}, 
while the B star seem to display, notably, a solar composition \citep{andrievsky95}.
If these abundances are confirmed, this could seriously defy the accretion scenario.
In addition, currently there is no analysis of the 3$^{rd}$ star, the late-type component of the system.
Therefore, we take the opportunity and perform a detailed abundance analysis including for the first time the three stars
of the system, using a spectra with higher resolving power than previous works.

We also present an analysis of the binary systems HD 193256/281 and HD 198160/161.
Both systems show solar values for C and subsolar Fe, similar to other candidate $\lambda$ Boo stars \citep{sturenburg93}.
However, more recent classification spectra suggest that only one star of the system belong to the $\lambda$ Boo
class \citep[see Tables 1 and 4 of ][]{murphy15,gray17},
which would be difficult to explain under the accretion scenario.
This convert both systems in very interesting targets to study in detail, and are included in our analysis.

This work is organized as follows. In Sect. 2 we describe the observations and data reduction, while in Sect. 3
we present the stellar parameters and chemical abundance \mbox{analysis}. In Sect. 4 we show the results and discussion,
and finally in Sect. 5 we highlight our main conclusions.

\section{Observations}

We present in Table \ref{table.parallax} the visual magnitude V (from Hipparcos), coordinates, proper motions
and parallax \citep[from Gaia DR2, ][]{gaiaDR2} for the stars studied in this work.
Spectral data of the triple system HD 15165 were obtained with the Max Planck Gesselschaft (MPG) 2.2 meter telescope
at the European Southern Observatory (ESO) in La Silla, Chile, on October 10, 2021 (Program ID: 0108.A-9012, PI: Marcelo Jaque Arancibia).
We used the Fiber-fed Extended Range Optical Spectrograph (FEROS), which provides a high-resolution (R$\sim$48.000) spectra
when illuminated via the 2.0 arcsec aperture on the sky in the unbinned mode.
Three individual spectra for each object were obtained, followed by a ThAr lamp in order to obtain an appropriate wavelength solution.
The data were reduced using the FEROS Data Reduction System\footnote{https://www.eso.org/sci/facilities/lasilla/instruments/feros/tools/DRS.html} (DRS).
The spectral coverage resulted between 3700-9000 \AA, approximately, and the S/N per pixel measured at $\sim$5000 \AA~resulted in $\sim$300.

\begin{table*}
\centering
\caption{Magnitudes and astrometric data for the stars studied in this work.}
\begin{tabular}{lccccccc}
\hline
Star        & V     & $\alpha$    & $\delta$     & $\mu_{\alpha}$ & $\mu_{\delta}$ & $\pi$   & Spectra  \\
            &       & J2000       & J2000        & [mas/yr]       & [mas/yr]       & [mas]   &          \\
\hline
HD 15164    &  8.27 & 02 26 48.29 & +10 34 57.59 & 36.552         & -13.717        &  7.4185 & MPG+FEROS    \\
HD 15165    &  6.69 & 02 26 45.65 & +10 33 55.07 & 36.680         & -13.086        &  7.4414 & MPG+FEROS    \\ 
HD 15165C   & 11.78 & 02 26 47.40 & +10 32 58.89 & 36.805         & -13.131        &  7.5499 & MPG+FEROS    \\ 
HD 193256   &  7.53 & 20 20 26.57 & -29 11 28.76 & -1.991         & -1.221         &  5.8675 & CASLEO+REOSC    \\ 
HD 193281   &  6.64 & 20 20 27.88 & -29 11 49.97 & -0.653         & 0.244          &  6.2644 & CASLEO+REOSC    \\ 
HD 198160   &  6.21 & 20 51 38.51 & -62 25 45.59 & 82.697         & -46.562        & 13.5137 & MPG+FEROS    \\ 
HD 198161   &  6.56 & 20 51 38.85 & -62 25 45.26 & 82.077         & -42.340        & 13.5315 & MPG+FEROS    \\ 
\hline
\end{tabular}
\normalsize
\label{table.parallax}
\end{table*}

The spectra of the binary system HD 193256/281 were obtained at the Complejo Astr\'onomico El Leoncito (CASLEO) 
between May 9 and 11, 2009 (PI: Maria Eugenia Veramendi). We used the \emph{Jorge Sahade} 2.15-m telescope equipped with a REOSC echelle
spectrograph\footnote{On loan from the Institute d'Astrophysique de Liege, Belgium.} and a TEK 1024$\times$1024 CCD detector.
The REOSC spectrograph uses gratings as cross dispersers. We used a grating with 400 lines mm$^{-1}$, which provides
a resolving power of $\sim$ 12500 covering the spectral range $\lambda\lambda$3800--6500.
Three individual spectra for each object were obtained and then combined, reaching a final S/N per pixel of $\sim$300 measured at $\sim$5000 \AA.
The data were reduced with Image Reduction and Analysis Facility (IRAF) following the standard recipe for echelle spectra (i.e. bias and flat corrections, 
order-by-order normalization, scattered light correction, etc.).

Finally, the FEROS spectra of the binary system HD 198160/161 were obtained from the ESO Science Archive 
Facility\footnote{http://archive.eso.org/cms.html}. The stars were observed between April 4 and 7, 2017 (Program ID: 099-A-9029).
The spectra were reduced using the FEROS DRS, obtaining a spectral coverage and S/N similar to those obtained with HD 15165.

\section{Stellar parameters and abundance analysis}

The stellar parameters T$_\mathrm{eff}$ and $\log g$ were estimated iteratively,
similar to previous works \citep{saffe21}.
They were first estimated by using the Str\"omgren uvby$\beta$ mean photometry of \citet{hauck-mermilliod98}
or considering previously published results.
We used the calibration of \citet{napi93} and deredenned colors according to \citet{domingo-figueras99}
within the program TempLogG \citep{kaiser06}, in order to derive the fundamental parameters.
These initial values were refined (when necessary and/or possible) by imposing excitation and ionization balances of the iron lines.
A similar strategy was previously applied in the literature \citep[e.g., ][]{saffe-levato14,saffe21}.
The values derived in this way are listed in the Table \ref{table.params}, with an average dispersion of
$\sim$115 K and $\sim$0.13 dex for T$_{\rm eff}$ and {$\log g$}, respectively.

\begin{table*}
\centering
\caption{Fundamental parameters derived for the stars in this work.}
\begin{tabular}{lcccc}
\hline
Star & T$_{\rm eff}$ & $\log g$ & v$_\mathrm{micro}$ & $v\sin i$     \\
     &  [K]          &  [dex]   & [km s$^{-1}$]      & [km s$^{-1}$] \\
\hline
HD 15164       &   7150  $\pm$ 70   &   3.74 $\pm$  0.08  &   2.54 $\pm$  0.63  &   17.9 $\pm$  0.7   \\ 
HD 15165       &   6950  $\pm$ 139  &   3.80 $\pm$  0.19  &   2.21 $\pm$  0.55  &  125.7 $\pm$  5.4   \\ 
HD 15165C      &   4960  $\pm$ 51   &   4.40 $\pm$  0.03  &   0.46 $\pm$  0.07  &    2.4 $\pm$  0.3   \\ 
HD 193256      &   7780  $\pm$ 146  &   3.97 $\pm$  0.19  &   3.23 $\pm$  0.81  &  257.0 $\pm$  8.2   \\ 
HD 193281      &   8700  $\pm$ 140  &   3.60 $\pm$  0.15  &   2.99 $\pm$  0.75  &   91.5 $\pm$  3.9   \\ 
HD 198160      &   8010  $\pm$ 130  &   4.09 $\pm$  0.15  &   3.31 $\pm$  0.83  &  190.0 $\pm$  6.8   \\ 
HD 198161      &   8010  $\pm$ 130  &   4.09 $\pm$  0.15  &   3.31 $\pm$  0.83  &  185.0 $\pm$  7.2   \\ 
\hline
\end{tabular}
\normalsize
\label{table.params}
\end{table*}

Projected rotational velocities $v\sin i$ were estimated by fitting most \ion{Fe}{I} and \ion{Fe}{II} lines in the spectra. 
Synthetic spectra were calculated using the program SYNTHE \citep{kurucz-avrett81} together with ATLAS12 \citep{kurucz93} model atmospheres.
Microturbulence velocity v$_\mathrm{micro}$ was estimated as a function of T$_{\rm eff}$ following the formula of \citet{gebran14}, 
(valid for $\sim$6000 K $<$ T$_{\rm eff}$ $<$ $\sim$10000 K), except for the late-type star
HD 15165C for which we used the formula of \citet{ramirez13} for FGK stars.
We adopt for v$_\mathrm{micro}$ an uncertainty of $\sim$25 $\%$, as suggested by \citet{gebran14}.

Chemical abundances were determined iteratively by fitting a synthetic spectra using the program SYNTHE \citep{kurucz93}.
In the first step, we use an ATLAS12 model atmosphere calculated with solar abundances.
With the new abundance values, we derived a new model atmosphere and started the process again.
In each step, opacities were calculated for an arbitrary composition and v$_\mathrm{micro}$ using the opacity
sampling (OS) method, similar to previous works \citep{saffe20,saffe21}.
Possible differences originated from the use of opacities with solar-scaled composition instead of an arbitrary
composition, were recently estimated for solar-type stars \citep{saffe18,saffe19}.
If necessary, T$_{\rm eff}$ and $\log g$ were refined to achieve the balance of Fe I and Fe II lines.
In this way, abundances and parameters are consistently derived until reach the same input and output abundance values
\citep[for more details, see ][]{saffe21}.

\begin{figure}
\centering
\includegraphics[width=8cm]{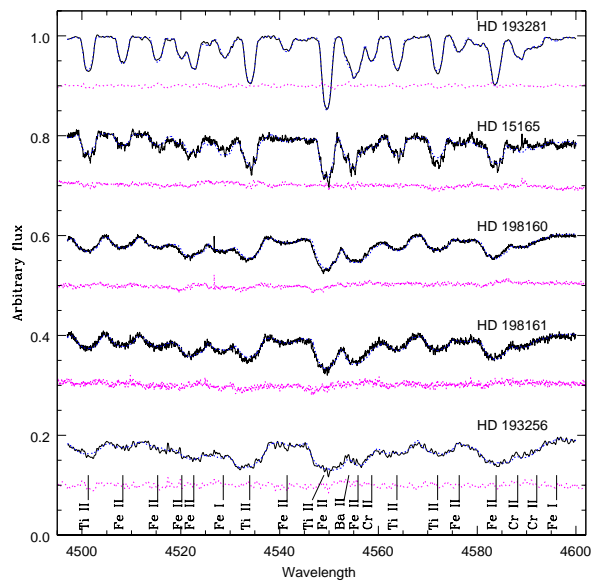}
\caption{Observed, synthetic, and difference spectra (black, blue dotted, and magenta lines) 
for the stars in our sample, sorted by $v\sin i$.}
\label{fig.region1}%
\end{figure}

\begin{figure}
\centering
\includegraphics[width=8cm]{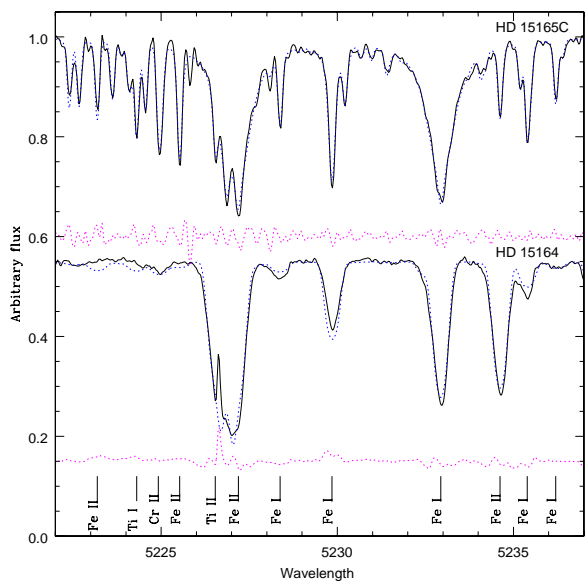}
\caption{Observed, synthetic, and difference spectra (black, blue dotted, and magenta lines) 
for the stars in our sample, sorted by $v\sin i$.}
\label{fig.region3}%
\end{figure}

Chemical abundances were derived for 24 different species. 
The atomic line list and laboratory data used in this work are the same described in \citet{saffe21}.
In Figs. \ref{fig.region1} and \ref{fig.region3} we present an example of observed and synthetic spectra
(black and blue dotted lines, almost superimposed) together with the difference spectra (magenta)
for the stars in our sample. For clarity, Fig. \ref{fig.region1} corresponds to stars with the higher $v\sin i$ values
($>$ 91 km s$^{-1}$), while Fig. \ref{fig.region3} corresponds to stars with the lower $v\sin i$ values ($<$ 17.9 km s$^{-1}$).
The stars are sorted in these plots by increasing $v\sin i$.
There is a good agreement between modeling and observations for the lines of different chemical species.
To determine the uncertainty in the abundance values, we considered different sources.
The total error e$_{tot}$ was derived as the quadratic sum of the line-to-line dispersion e$_{1}$
(estimated as $\sigma/\sqrt{n}$ , where $\sigma$ is the standard deviation),
and the error in the abundances (e$_{2}$, e$_{3}$ and e$_{4}$) when varying T$_{\rm eff}$, $\log g$
and v$_\mathrm{micro}$ by their corresponding uncertainties\footnote{We adopt a minimum of 0.01 dex for the errors e$_{2}$, e$_{3}$ and e$_{4}$.}.
For chemical species with only one line, we adopt as $\sigma$ the standard deviation of iron lines.
The abundances, the total error e$_{tot}$ and the individual errors e$_{1}$ to e$_{4}$ are presented in Tables 
B.1 to B.7 of the Appendix.

\subsection{NLTE effects}

Light element Non-Local Thermodynamic Equilibrium (NLTE)
abundances are particularly important for the case of $\lambda$ Boo stars.
For instance, \citet{paunzen99} derived for a sample of $\lambda$ Boo stars an average O I correction
of -0.5 dex, while for C I they estimated an average correction of -0.1 dex.
\citet{rentzsch96} calculated carbon NLTE abundance corrections by using a multilevel 
model atom for stars with T$_\mathrm{eff}$ between 7000 K and 12000 K, log g between
3.5 and 4.5 dex, and metallicities from -0.5 dex to +1.0 dex.
She showed that C I NLTE effects are negative (calculated as NLTE-LTE) and
depend basically on equivalent width W$_{eq}$.
Near $\sim$7000 K, the three lower levels of C I are always in LTE; however, by increasing
the T$_\mathrm{eff}$ increase the underpopulation of these levels respect to LTE
by UV photoionization.
Then, we estimated NLTE abundance corrections of C I for the early-type stars in our sample
by interpolating in their Figs. 7 and 8 as a function of T$_\mathrm{eff}$, W$_{eq}$
and metallicity.

\citet{sitnova13} performed NLTE abundance corrections for O I for stars with spectral
types from A to K (T$_\mathrm{eff}$ between 10000 and 5000 K).
They showed that NLTE effects lead to an strengthening of O I lines,
producing a negative NLTE correction.
We estimated NLTE abundance corrections of O I (IR triplet 7771 \AA\ and 6158 \AA)
for the stars in this work, by interpolating in the Table 11 of \citet{sitnova13} as
a function of T$_\mathrm{eff}$. Other O I lines present corrections lower than $\sim$-0.02 dex
\citep[see, e.g., Table 5 of ][]{sitnova13}.

\subsection{Comparison with literature}

\begin{figure}
\centering
\includegraphics[width=8.0cm]{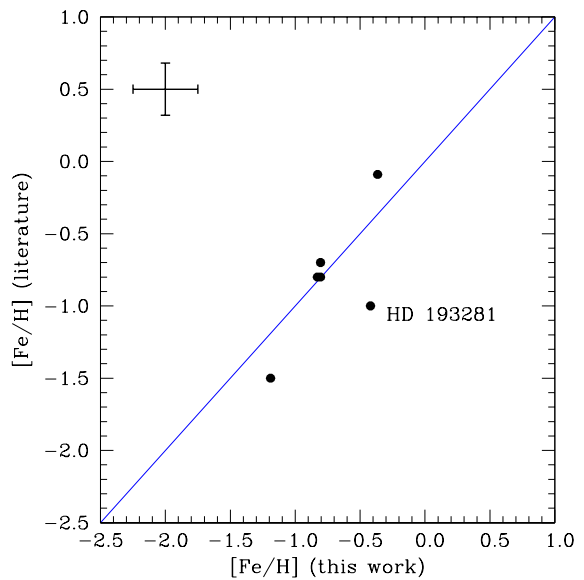}
\caption{Comparison of [Fe/H] values derived in this work with those from literature.
Average dispersion bars are showed in the upper left corner.}
\label{fig.metal.liter}
\end{figure}

We present in Fig. \ref{fig.metal.liter} a comparison of [Fe/H] values
derived in this work, with those taken from literature for the stars
HD 15164 \citep{andrievsky95}, HD 15164 \citep{paunzen02},
HD 193256, HD 193281, HD 198160 and HD 198161 \citep{sturenburg93}.
In general, there is a reasonable agreement with literature,
where the star HD 193281 present the larger difference (marked in the plot).

\citet{sturenburg93} estimated for HD 193281 an iron abundance of [Fe/H]=-1.0$\pm$0.2.
However, we estimated for this star a somewhat higher value of [FeI/H]=-0.36$\pm$0.13 ([FeII/H]=-0.48$\pm$0.13).
We explored the possible sources for this difference.
They estimated a T$_\mathrm{eff}$ of 8080 K (without quoting uncertainties) by using the Str\"omgren photometry, while
we estimated for this object a T$_\mathrm{eff}$ of 8700$\pm$140K, having a difference of 620 K.
This could be one of the reasons for the different [Fe/H] that we obtained.
Different works estimated for this star temperatures of 8700 K \citep{gray17}, 
8623 K \citep{koleva12}, and recently 8695 K \citep{arentsen19}.
Then, our estimated T$_\mathrm{eff}$ is more in agreement with these works.
We also note that this star presents different metallicities in literature:
-1.0$\pm$0.2 dex \citep{sturenburg93}, -0.68 dex \citep{koleva12} and more recently -0.37 dex \citep{arentsen19}.
Our estimated metallicity of [FeI/H]=-0.36$\pm$0.13 is closer to the work of \citet{arentsen19}.

In addition, there is evidence that HD 193281 could be contaminated by a nearby star.
Simbad database reports that the star ADS 13702 B (= TYC 6918-1823-2) is located at $\sim$3.5 arcsec
from HD 193281, having spectral type "F5:V".
\citet{ivanov19} present a library of stellar spectra taken with the integral field spectrograph
MUSE\footnote{https://www.eso.org/sci/facilities/develop/instruments/muse.html}
in low spectral resolution (R$\sim$2000) although with high spatial resolution (0.3-0.4 arcsec).
They report that HD 193281 is a binary with $\sim$3.8 arcsec separation and the
components cross-contaminate each other. They identified the components as HD 193281 A and B,
and estimated spectral types A2 III and K2 III, respectively (updating the spectral type
F5:V reported by Simbad for the star HD 193281 B).
This possible contamination could explain, at least in part, the different parameters and 
metallicities obtained from different works for this object.
In this study, we estimated parameters and abundances of HD 193281 taken as single,
for which the resulting values should then be considered with caution.

\section{Discussion}

In order to test the accretion scenario of $\lambda$ Boo stars, we compare the chemical abundances
of the stars in our sample with those of $\lambda$ Boo stars.
The three multiple systems with candidate $\lambda$ Boo stars are discussed
separately, while other binary or multiple systems with candidate $\lambda$ Boo stars
are discussed in the Appendix.

\subsection{The average pattern of $\lambda$ Boo stars}

To derive an average $\lambda$ Boo pattern is not an easy task.
Few literature works obtain homogeneous abundances of many species for $\lambda$ Boo stars \citep[e.g., ][]{sturenburg93,andrievsky02,heiter02}.
\citet{sturenburg93} derived abundances for 16 A-type stars classified, in principle, as $\lambda$ Boo stars.
They performed NLTE corrections for some elements including C.
However, they included stars that were subsequently considered non-members or uncertain members, such as HD 38545 
and HD 193281 \citep{murphy15}.
\citet{paunzen99} and \citet{kamp01} derived light-element NLTE abundances for a sample of $\lambda$ Boo stars.
Then, \citet{andrievsky02} derived elemental abundances for 20 candidate $\lambda$ Boo stars basically selected from
classification-resolution spectroscopy. They performed primarily a LTE approach and included NLTE effects for Na.
They were able to confirm the membership of only nine objects to the $\lambda$ Boo class,
while other stars were ruled out or present an unclear membership.
\citet{paunzen02} collected abundance values for 26 candidate $\lambda$ Boo stars (see their Table 5),
although using different literature sources.
Also, \citet{heiter02} reported LTE abundance values for 12 candidate $\lambda$ Boo stars, four of them belonging
to SB systems.
Then, it would be highly desirable a homogeneous abundance determination including more candidate 
$\lambda$ Boo stars, newer laboratory data for the lines and including NLTE effects especially for the light-elements.

In order to test the accretion scenario of $\lambda$ Boo stars, we compare the chemical abundances
of the stars in our sample with those of $\lambda$ Boo stars.
In this work, we used the data derived by \citet{heiter02}, who homogeneously determined
abundances for a number of $\lambda$ Boo stars.
We excluded from the average those stars without CNO values and the stars analyzed here.

\subsection{The triple system HD 15164/65/65C}

This remarkable triple system is composed by two early-type stars (HD 15165 and HD 15164, the stars A and B)
and a late-type companion (HD 15165C).
A number of studies suggest that the spectrum of HD 15165 resembles that of metal-deficient star,
but the companion HD 15164 has a near solar abundance \citep{mechler74,mechler76,abt80}.
Then, as explained in the Introduction, some works suggest that the A star belong to the $\lambda$ Boo class \citep{andrievsky95,cherny98}, 
while the B star seems to display a solar composition \citep{andrievsky95}.
To our knowledge, there is no abundance determination for the C component.

\begin{figure}
\centering
\includegraphics[width=8.0cm]{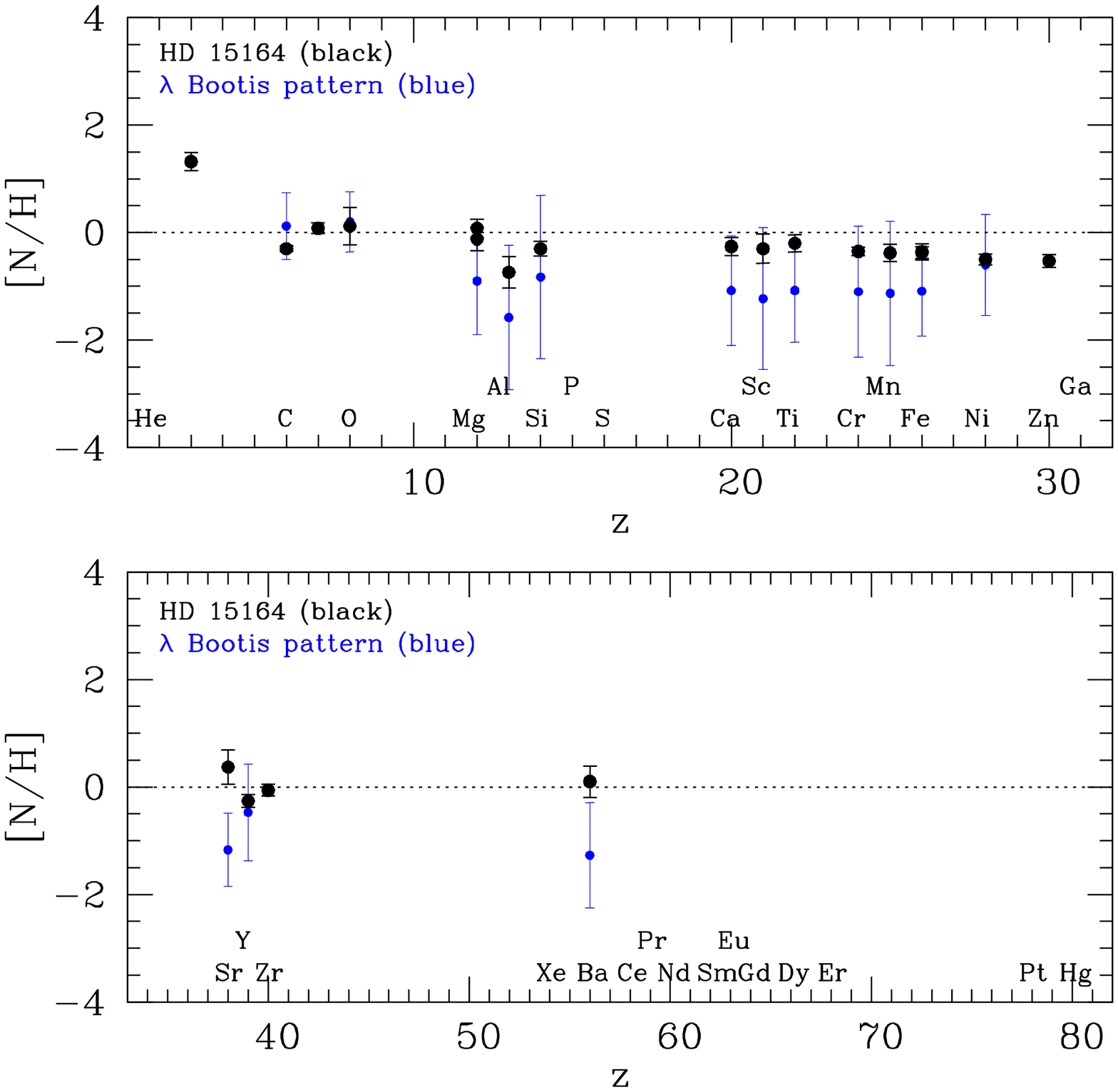}
\includegraphics[width=8.0cm]{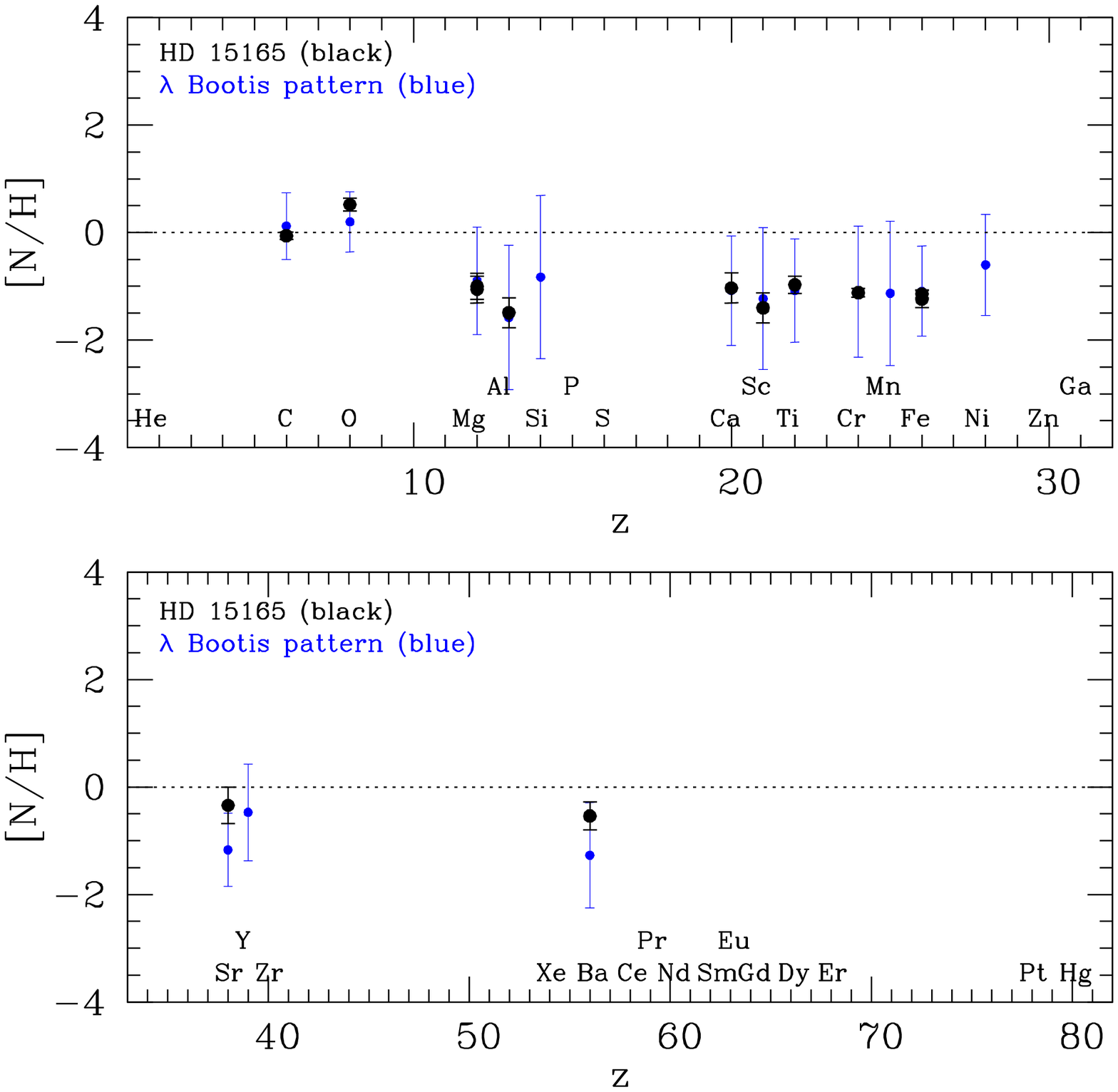}
\includegraphics[width=8.0cm]{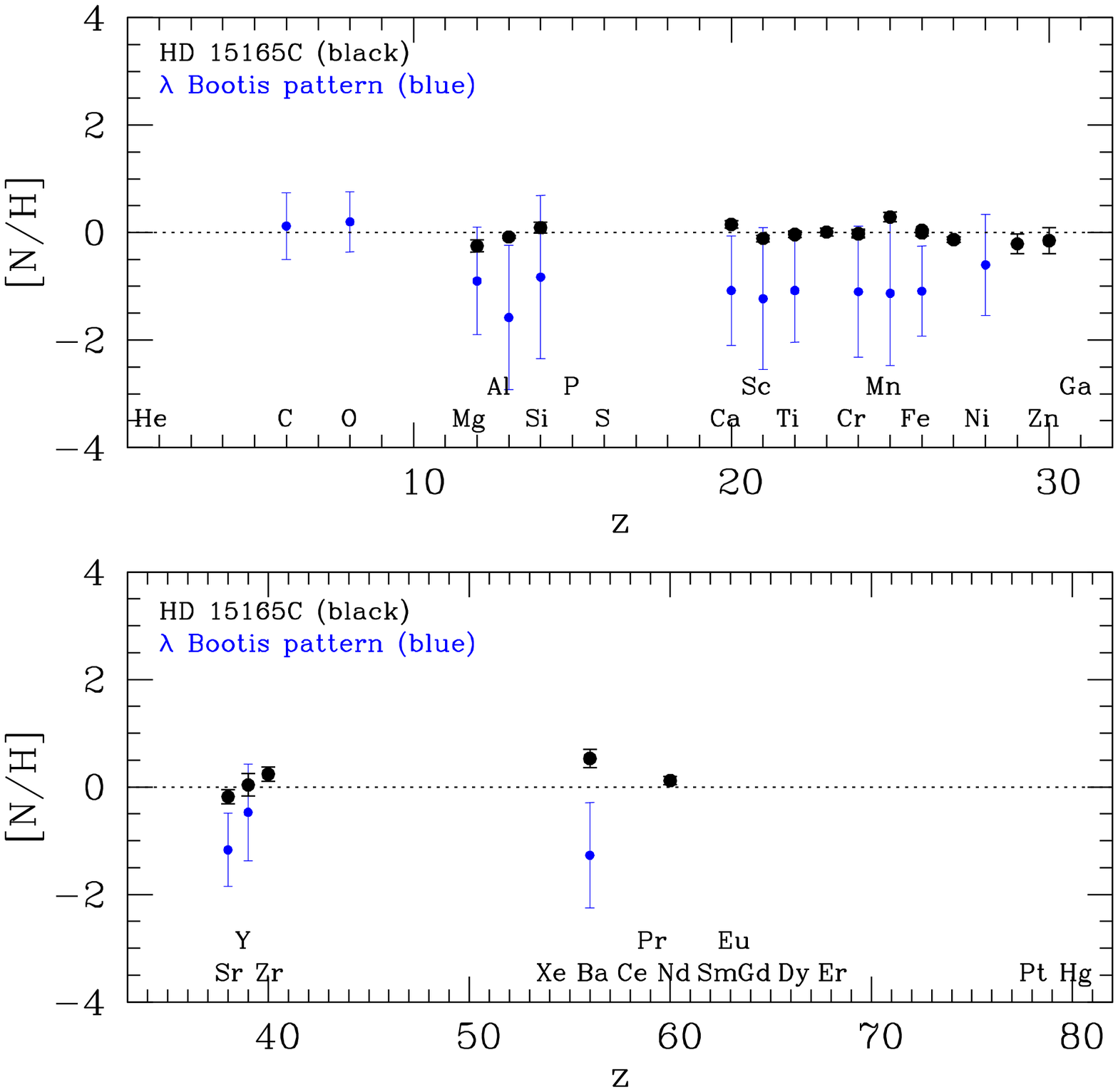}
\caption{Chemical pattern of the stars HD 15164, HD 15165 and HD 15165C (black), 
compared to an average pattern of $\lambda$ Boo stars (blue).}
\label{fig.pattern.HD15165}
\end{figure}

We present in Fig. \ref{fig.pattern.HD15165} the chemical pattern of the stars
HD 15164, HD 15165 and HD 15165C (black), compared to an average pattern of
$\lambda$ Boo stars (blue). For each star we present two panels, 
corresponding to elements with atomic number z$<$32 and z$>$32.
The error bars of the $\lambda$ Boo pattern show the standard deviation derived from different stars,
while the error bars for our stars correspond to the total error e$_{tot}$.
As we can see in the Fig. \ref{fig.pattern.HD15165} the chemical pattern of the primary (HD 15165) 
is similar to the pattern of $\lambda$ Boo stars, showing subsolar abundances of most metals
(Mg, Al, Ca, Sc, Ti, Cr, Fe) together with near solar values of C and O.
The abundances of Sr and Ba present a less marked deficiency, although still showing subsolar values.
On the other hand, the chemical pattern of the secondary star (HD 15164) shows a slight deficiency
in some metals (for instance [Fe/H]=-0.36$\pm$0.15 dex), although closer in general to the solar pattern 
than to the $\lambda$ Boo stars. In this sense, a primary showing a $\lambda$ Boo pattern
and a secondary showing near solar abundances, verify the early result of \citet{andrievsky95}:
the early-type stars A and B present different chemical compositions.

To our knowledge, there is no abundance determination of $\lambda$ Boo stars that belong
to a triple or multiple system. In particular, a late-type star that belong to such system,
could be used as a proxy of the initial composition of the material where the $\lambda$ Boo star formed
(under the hypothesis that they born from the same molecular cloud).
This could be important as an additional constrain for any model trying to explain the $\lambda$ Boo phenomena.
We present in the Fig. \ref{fig.pattern.HD15165} the chemical pattern of HD 15165C,
the late-type component of the triple system.
The chemical pattern is compatible with a solar-like composition (for instance, [FeI/H]=0.04$\pm$0.02 dex).
This is in agreement with the idea that $\lambda$ Boo stars are Population I objects
and originate (following any internal or external mechanism) starting from a solar-like composition.

\begin{figure}
\centering
\includegraphics[width=8.0cm]{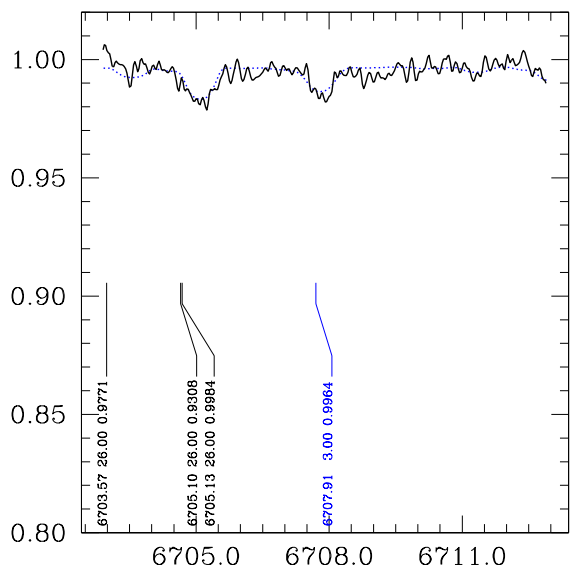}
\caption{Observed spectra (black line) and synthetic spectra (blue dotted line) near the Li line 6707.8 \AA\ in the star HD 15164.
Synthetic lines are indicated showing the wavelength, atomic number and intensity.}
\label{fig.HD15164.Li.6707}
\end{figure}

Notably, the three stars that belong to the triple system present different chemical patterns.
The star A present a $\lambda$ Boo pattern, while the stars B and C present abundances closer
to the Sun. However, the stars B and C are also slightly different between them:
the late-type star C present the closest abundances to the Sun, while the early-type star B
shows a slightly deficiency. Most abundance values between stars B and C agree within $\sim$0.30 dex,
with a possible exception: the lithium content.
The Li I 6707.8 \AA\ line is clearly present in the spectra of the star B (HD 15164) as we can see in the Fig. \ref{fig.HD15164.Li.6707},
while it is not detected in the spectra of stars A nor C.
It is interesting to note that this line is commonly used as a proxy of recent accretion onto the 
atmosphere of the stars. For instance, \citet{saffe17} attributed a notable difference in the refractory
abundances and in the Li content between the stars of the binary system HAT-P-4
to a possible accretion event of a rocky planet onto the primary.
However, although HD 15164 shows clearly the Li line, its refractory content is slightly lower than
the star HD 15165C, which would be difficult to explain with the accretion of refractory species.

Is it possible that the supposed different abundances between stars A, B and C are only due to different T$_\mathrm{eff}$?
The question makes sense because the stars A and C present T$_\mathrm{eff}$ of 7150 K and 4960 K,
a difference of 2190 K. However, the total error e$_{tot}$ in abundances includes the error e$_{2}$, which
measure the change in the abundances when varying T$_\mathrm{eff}$ by their corresponding uncertainty.
Then, we do not expect a strong change in the derived abundances due to T$_\mathrm{eff}$
(in any case, the possible change is contained within the total error e$_{tot}$).

\subsection{The binary system HD 193256/281}

HD 193256 was classified as $\lambda$ Boo by \citet{gray88} and then as uncertain $\lambda$ Boo
by \citet{renson90}. It is separated by $\sim$27.5 arcsec from HD 193281,
which was classified as $\lambda$ Boo by \citet{gray-garrison87}.
Both stars HD 193256 and HD 193281 show approximately solar abundances of C and subsolar Fe
in the study of \citet{sturenburg93}, who analyzed them separately.
However, they also found near solar values for other elements such as Mg and Si in both stars,
which is different from what found in average $\lambda$ Boo stars.
\citet{kamp01} found solar values in HD 193281 for N, O and S, although for C they found -0.61 dex,
similarly to \citet{paunzen99}.
However, more recent classification spectra suggest that only HD 193256 could belong to the $\lambda$ Boo class
\citep[see Tables 1 and 4 of ][]{murphy15,gray17}, while HD 193281 display a normal spectra.

\begin{figure}
\centering
\includegraphics[width=8.0cm]{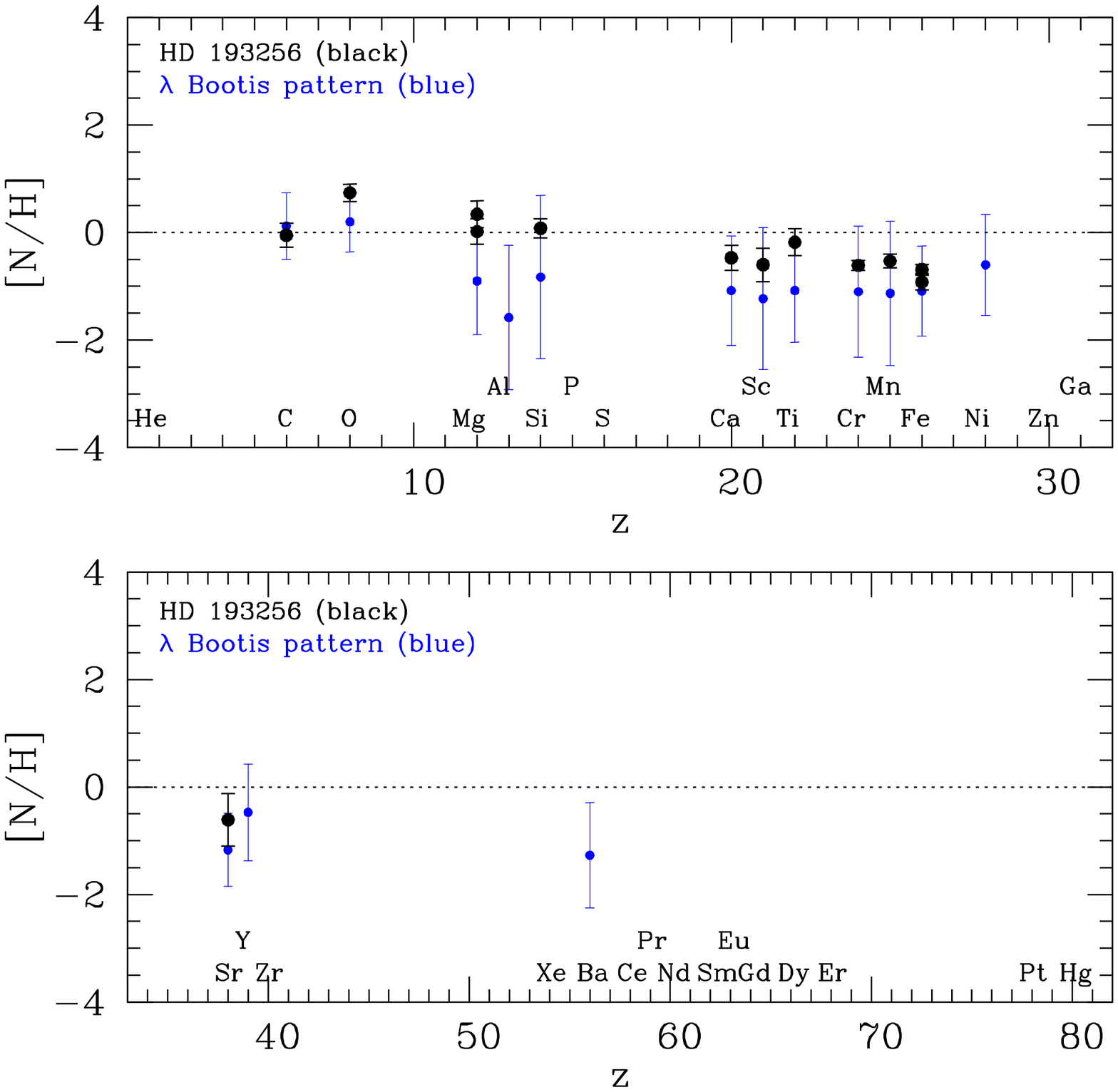}
\includegraphics[width=8.0cm]{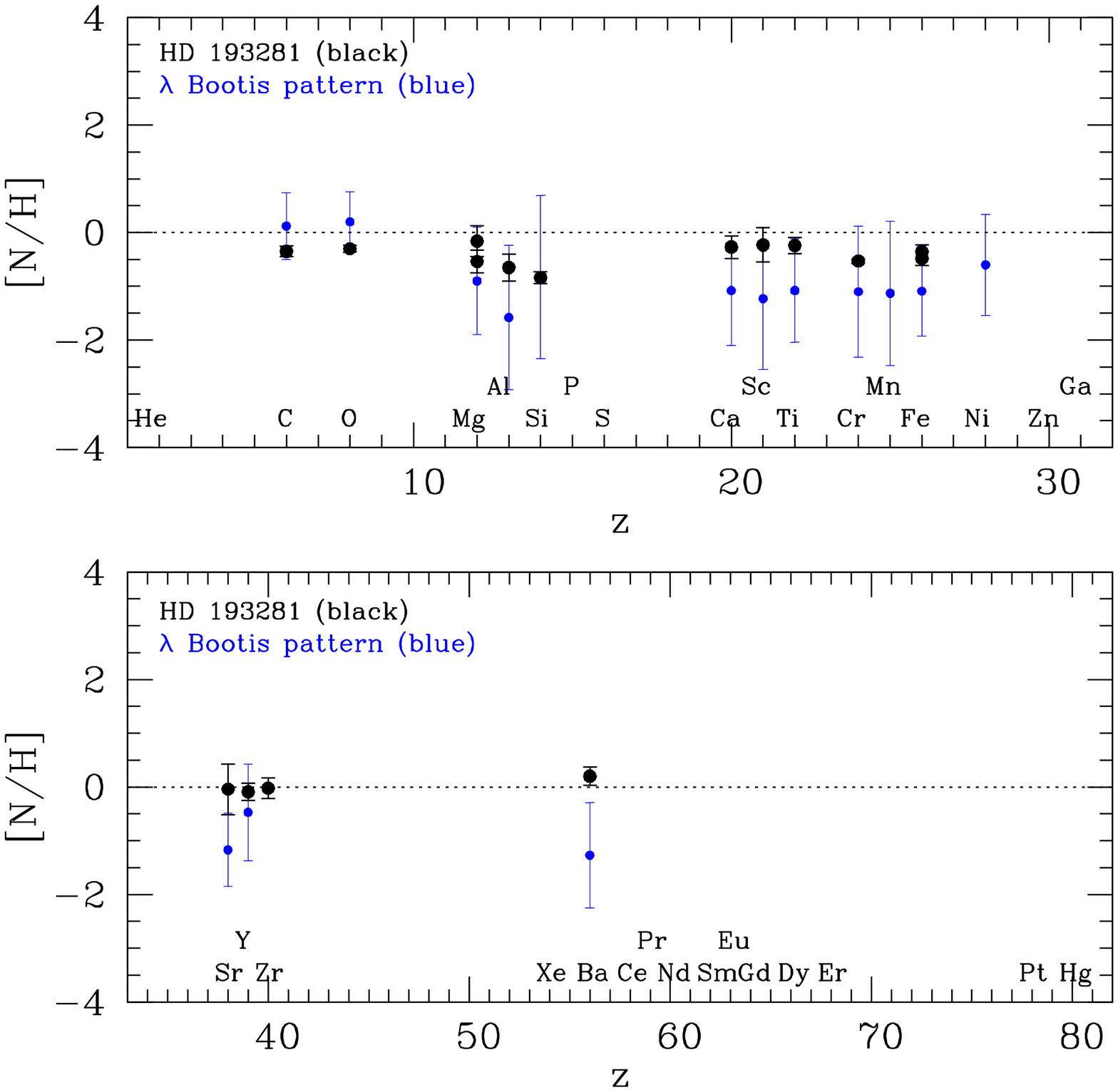}
\caption{Chemical pattern of the stars HD 193256 and HD 193281 (black), 
compared to an average pattern of $\lambda$ Boo stars (blue).}
\label{fig.pattern.HD193256}
\end{figure}

In this work, we analyzed the spectra of HD 193256 and HD 193281 considered both as single,
for which the abundances of HD 193281 should be taken with caution.
We present in Fig. \ref{fig.pattern.HD193256} the chemical pattern of the stars
HD 193256 and HD 193281 (black), compared to an average pattern of $\lambda$ Boo stars (blue).
The colors, panels and error bars used are similar to those of Fig. \ref{fig.pattern.HD15165}.
HD 193256 shows solar or suprasolar values for C and O, together with subsolar values (between 0.5-0.9 dex)
of Ca, Cr, Fe and Sr.
However, we also found near solar values of Mg, Si and Ti, which is not common in $\lambda$ Boo stars.
Then, this object seem to present a mix of metals with solar and subsolar abundances.
On the other hand, HD 193281 present the chemical pattern of a slightly metal-deficient star
in general, showing subsolar values for C and O ($\sim$0.3 dex) similar to Fe I (-0.36$\sim$0.13 dex).
However, the results of HD 193281 should be taken with caution, due to a possible contamination
of the nearby K2 III star.

In short, the solar abundances of some metals of HD 193256 (Mg, Si and Ti) are different of $\lambda$ Boo stars.
The chemical pattern of HD 193281 (considered as single) shows a slightly metal deficient star.
In addition, there is evidence for a possible contamination of HD 193281, where the components A and B
display spectral types A2 III and K2 III.
Then, current evidence does not support the presence of two bonafide $\lambda$ Boo stars in this binary (or triple) system.
It would be desirable an analysis of HD 193281 separately for the components A and B, 
in order to more properly determine the individual abundances.

\subsection{The binary system HD 198160/161}

HD 198160 form a visual binary system with HD 198161, separated by $\sim$2.4 arcsec.
HD 198160 was classified "A2 Vann wk4481" and "A2 Vn" \citep{gray88,corbally-garrison80},
while HD 198161 was classified as "A3 Vn" \citep{corbally-garrison80}.
Both stars were studied separately by \citet{sturenburg93} considering them as twins (same T$_\mathrm{eff}$ and log g).
He derived near solar values for C in both stars and subsolar values for
Fe (-0.8$\pm$0.2 dex), however he also obtained solar values for Mg and Si (0.0$\pm$0.1 dex 
and -0.2$\pm$0.2 dex for both stars).
Then, \citet{paunzen99} estimated near solar NLTE values for C and O, although quoted 
for HD 198160/1 (not separated). More recently, \citet{murphy15} caution that individual NLTE
volatile abundances for HD 198160 and HD 198161 are not confirmed (such as those reported in this work)
and tentatively adopt for HD 198160 a classification "A2 Vann $\lambda$ Boo". However, its companion HD 198161
was classified as a normal star, with spectral type "A3 V" and "A3 IV(n)" \citep{murphy15,gray17}.

\begin{figure}
\centering
\includegraphics[width=8.0cm]{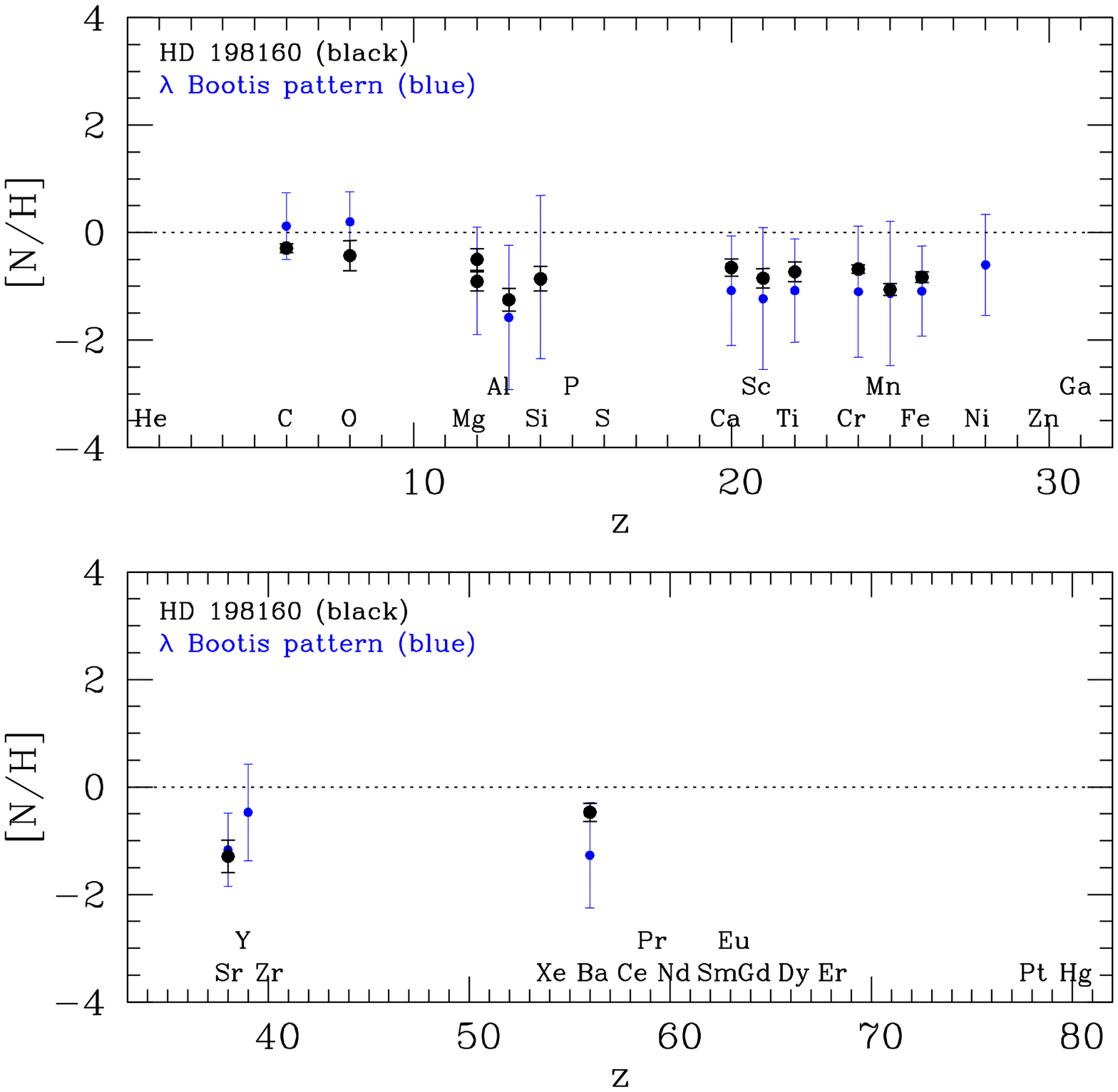}
\includegraphics[width=8.0cm]{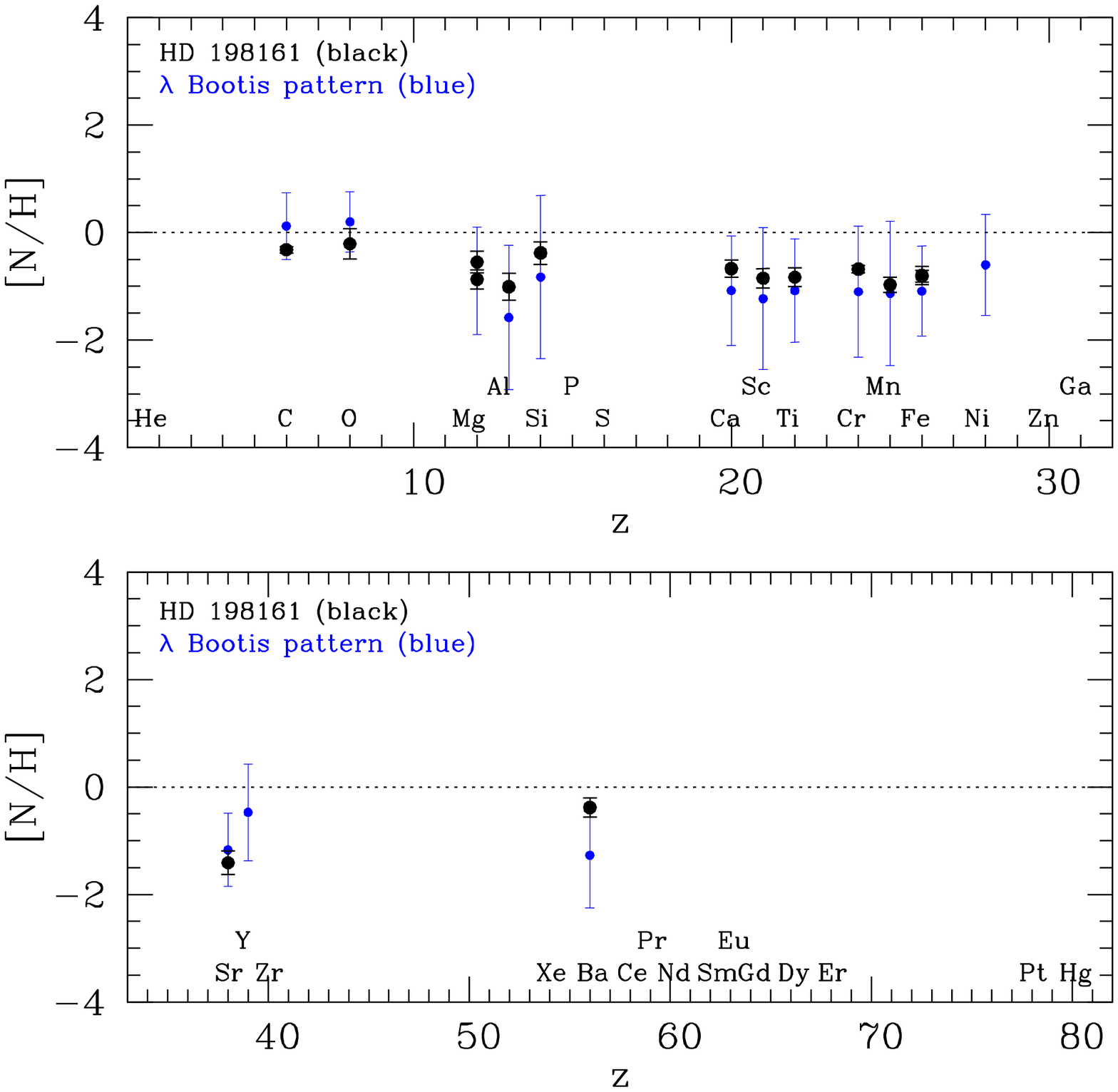}
\caption{Chemical pattern of the stars HD 198160 and HD 198161 (black), 
compared to an average pattern of $\lambda$ Boo stars (blue).}
\label{fig.pattern.HD198160}
\end{figure}

We present in Fig. \ref{fig.pattern.HD198160} the chemical pattern of the stars
HD 198160 and HD 198161 (black), compared to an average pattern of $\lambda$ Boo stars (blue).
The colors, panels and error bars used are similar to those of Fig. \ref{fig.pattern.HD15165}.
In both stars, most Fe-peak metals show a deficiency around 0.7-0.8 dex, similar to $\lambda$ Boo stars.
However, C and O also show subsolar values, being possibly low compared to other $\lambda$ Boo stars.
When comparing C with Fe abundances, the group of $\lambda$ Boo stars present [C/Fe]$\sim$1.21$\pm$0.35 dex \citep[excluding stars without CNO values
and the stars analyzed here, ][]{heiter02} with minimum and maximum values of 0.70 and 1.74 dex. 
However, the stars HD 198160 and HD 198161 present [C/Fe] values of $\sim$0.54 and $\sim$0.48 dex,
being low compared to the average [C/Fe] and even lower than the minimum of 0.70 dex.
Then, we consider that these low [C/Fe] values possibly correspond to mild-$\lambda$ Boo stars,
rather than to an average $\lambda$ Boo object.
It is important to note that our C and O abundances were corrected by NLTE,
with average corrections of -0.15 dex and -0.81 dex for both stars.
In other words, if we only adopt LTE values without correction, 
the C and O abundances would result closer to those of $\lambda$ Boo stars.

\subsection{On the physical association of the stars}

The stars studied in this work were previously reported as (possible) members
of binary/multiple systems, for the case of
HD 15164/65/65C \citep{andrievsky95,cherny98,murphy15},
HD 193256/281 \citep{paunzen12a,murphy15,gray17} and
HD 198160/161 \citep{paunzen12a,murphy15}.

The coordinates, proper motions and parallax of the stars (see Table \ref{table.parallax})
suggest that they are, at least, common proper motion objects.
We searched our targets stars in different binary catalogs from literature \citep{shaya-olling11,tokovinin-lepine12,andrews17}.
In particular, \citet{andrews17} performed a search of binaries through a Bayesian formulation
in the Tycho-Gaia catalogs and derived likelihoods of Keplerian orbits.
For HD 15164/65, they reported a probability greater than 99\% that they form
a physical system.
\citet{shaya-olling11} developed a Bayesian method to discover non-random pairs using Hipparcos data.
They include HD 198160/161 in their catalog of binaries, however there is no 
probability quoted for this pair.
Finally, we find no record for HD 193256/181 in these binary catalogs.

In this work, we assume that the stars form physical binary/multiple systems.
In the case of showing that the stars are not gravitationally bound,
then these stars would not be useful to test the accretion scenario.

\subsection{Are there two bonafide $\lambda$ Boo stars in binary systems?}

There is evidence in the literature supporting the accretion scenario.
For example, the anticorrelation of C and O with Si \citep{paunzen99}, first noted by \citet{hs93} for C.
It is expected that refractory elements like Fe and Si are condensed in dust, while the more volatile CNO and S
remain in the gaseous phase. Then, the selective accretion of gas will produce ratios [C/Si] or [O/Si]
larger than solar and reduced metallicity \citep{paunzen99}. 
\citet{kamp01} reached a similar conclusion comparing the volatile species N and S with the more refractory Ca.
We should also expect that in stars with large $v\sin i$, the meridional circulation mixes material of
solar composition from the stellar interior into the convection zone so that any surface contamination due to
accretion of circumstellar material should vanish. This observation seems to be weakly verified \citep[see e.g., ][]{solano01},
and would require a larger sample of $\lambda$ Boo stars.
As we can see, the accretion scenario could be tested by different methods.

In this work, we focus on the presence of $\lambda$ Boo stars as members of binary systems
\citep[e.g., ][]{sturenburg93,paunzen02,heiter02,paunzen12a,paunzen12b}.
These are the following 12 systems (see Appendix):
HD 15164/65/65C, HD 38545, HD 64491, HD 84948, HD 111786, HD 141851, HD 148628/638,
HD 171948, HD 174005, HD 193256/281, HD 198160/161, and HD 210111.
Following the accretion scenario, two early-type stars in a binary system should display, in principle,
a similar $\lambda$ Boo pattern after passing through a diffuse cloud.
However, a binary or multiple system having a $\lambda$ Boo star together with a "normal" early-type component
would be difficult to explain under the accretion scenario.
This test of the accretion scenario would require a detailed analysis of both stars.
As explained in the Introduction, some stars that belong to these 12 systems were recently classified
as non-members or uncertain members of the $\lambda$ Boo class, such as HD 141851, HD 148638 and HD 193256
\citep[see, e.g., ][]{murphy15,gray17}.
Then, we wonder if any of these 12 systems really include two stars with bonafide $\lambda$ Boo chemical patterns.

It would be desirable a detailed abundance analysis in order to verify the true $\lambda$ Boo nature of a star,
initially suggested (for instance) by its classification spectra \citep[see, e.g., ][]{andrievsky02,heiter02}.
To our knowledge, only 5 out of the 12 systems present an abundance determination of both components:
HD 15164/65, HD 84948, HD 171948, HD 193256/281 and HD 198160/161 (three of them were analyzed in this work).
Some works present an abundance study only of the brighter component, such as in the case of HD 38545 \citep{sturenburg93}
or HD 64491 \citep{kamp01}, while other systems only have a spectral classification, such as HD 174005 \citep{gray17,murphy15}.

An inspection of the abundance values reported in the literature (see Appendix) shows that, in our opinion, 
there is no binary system having two stars with bonafide $\lambda$ Boo chemical patterns.
The same is valid for the three systems analyzed in this work (HD 1564/65/65C, HD 193256/281 and HD 198160/161).
In fact, we cannot find even one binary system where the two stars present bonafide $\lambda$ Boo abundance patterns.
We consider that the closer candidates to show both stars a $\lambda$ Boo pattern are possibly the binary systems
HD 84948, HD 171948 and HD 198160. These three systems show [C/Fe] values lower than 0.7 dex (the minimum [C/Fe]
of $\lambda$ Boo stars, see Sect. 4.4 and Appendix),  
being perhaps mild-$\lambda$ Boo systems rather than clear $\lambda$ Boo objects. 
Then, we find no clear evidence for the presence of two $\lambda$ Boo stars as members of binary systems.
However, this fact (if confirmed) do not rule out the accretion scenario.

On the other hand, a challenge for the accretion scenario, would be the presence of a bonafide $\lambda$ Boo star
and a normal early-type object, together in the same multiple system.
By reviewing the 12 systems studied (including the stars of this work) we found only one candidate: the system HD 15164/65/65C
analyzed here. The star A present a $\lambda$ Boo pattern, while the stars B (early-type) and C (late-type) present
abundances closer to the Sun. 
The different chemical composition between stars A and B was initially
attributed to a possible stellar capture \citep{andrievsky95}.
The probability of a binary capture depends on several factors, such as the number of stars per cubic parsec, the velocity
dispersion and the mass of the stars \citep[e.g.,][]{clarke-pringle91,boffin98}.
The capture is not a dominant formation process for solar-mass (coeval) binaries in dense
clusters \citep[e.g.,][]{clarke-pringle91,heller95,boffin98}.
To our knowledge, there is no known binary or triple system with an origin attributed to a capture.
On the other hand, there are multiple observations of young binaries embedded in dense cores \citep[e.g., ][]{sadavoy-stahler17},
and even an image of a triple protostar formed via disk fragmentation \citep{tobin16}.
Although the capture cannot be totally discarded, most observational evidence points toward the formation
of binary and multiple systems from a common molecular cloud.
Taking up the idea that the three stars are born together, it is difficult to explain the composition
of the stars of HD 15165 under the accretion scenario.
Then, there is an urgent need of additional binary and multiple systems to be analyzed through a detailed
abundance analysis, in order to test the accretion model of $\lambda$ Boo stars.

\section{Concluding remarks}

In the present work, we performed a detailed abundance determination of selected binary and multiple systems
with candidate $\lambda$ Boo stars, in order to test the accretion scenario.
Reviewing abundance values reported in the literature (see Appendix) shows that, in our opinion, there are no
binary system having two stars with bonafide $\lambda$ Boo chemical patterns. The same is valid for the three systems
analyzed in this work (HD 15164/65/65C, HD 193256/281 and HD 198160/161).
We consider that the closer candidates to show both stars a $\lambda$ Boo pattern are possibly the binary systems
HD 84948, HD 171948 and HD 198160. However, these three binary systems are perhaps mild-$\lambda$ Boo systems rather
than clear $\lambda$ Boo objects. Then, in our opinion, current evidence of binary/multiple systems does not give
strong support to the accretion scenario of $\lambda$ Boo stars.

On the other hand, a binary/multiple system formed by a $\lambda$ Boo star and an early-type "normal" object,
would be difficult to explain under the accretion scenario. We found one candidate: the remarkable triple system HD 15164/65/65C. 
It is composed by two early-type stars (A and B) and a late-type companion (C). 
In particular, the late-type component of the system could be used as a proxy for
the initial composition of the system, constraining formation models of $\lambda$ Boo stars.
We found a $\lambda$ Boo pattern for the A star (HD 15165), while the stars B and C present
abundances closer to the Sun. Then, there is an urgent need of additional binary and multiple systems 
to be analyzed through a detailed abundance analysis, in order to test the accretion model of $\lambda$ Boo stars.

\begin{acknowledgements}
We thank the referee Dr. Christopher Corbally for constructive comments that improved the paper.
The authors thank Dr. R. Kurucz for making their codes available to us.
CS acknowledge financial support from FONCyT (Argentina) through grant PICT 2017-2294
and the National University of San Juan (Argentina) through grant CICITCA E1134.
IRAF is distributed by the National Optical Astronomical Observatories, 
which is operated by the Association of Universities for Research in Astronomy, Inc., under a cooperative agreement
with the National Science Foundation.
Based on data acquired at Complejo Astron\'omico El Leoncito, operated under agreement between
the Consejo Nacional de Investigaciones Cient\'ificas y T\'ecnicas de la Rep\'ublica Argentina and
the National Universities of La Plata, C\'ordoba and San Juan.

\end{acknowledgements}

\begin{appendix}

\section{Multiple systems with suspected $\lambda$ Boo components}

We review abundance determination of binary or multiple systems with suspected $\lambda$ Boo components
from the literature, in order to determine if two bonafide $\lambda$ Boo stars can be found.
Spectral classification data is also included whenever available.
The data are updated including the results from the present work.

{\bf{HD 15164/65/65C}}:
It is a visual triple system, where most works considered only the two brighter components.
\citet{andrievsky95} studied spectra of the stars A and B (HD 15165 and HD 15164) using the LYNX (R$\sim$24000)
and AURELIE (R$\sim$11000) spectrographs.
They found subsolar values for two elements analyzed in the A star (-0.73 dex for [Ca/H] and -0.46 dex for [Fe/H]).
Then, \citet{cherny98} reanalyzed the data for the A star and suggest that this object belongs to the $\lambda$ Boo class,
showing $\sim$solar values for C, O and S (0.0 dex, -0.3 dex and 0.0 dex) together with subsolar values for 
refractory elements (for example, [Fe/H]=-1.6 dex).
However, \citet{andrievsky95} also found solar values for several elements in the B star. They suggest that
the different chemical composition of stars A and B is probably due to a stellar capture.
\citet{murphy15} classified the spectra of the 3 stars as "F1 V kA7mA6 ($\lambda$ Boo)?" (HD 15164), 
"F2 V kA2mA2 $\lambda$ Boo?" (HD 15165) and "K2V" (HD15165C). They also claim that the classification spectrum
of HD 15165 does not match solar abundances, contrary to the result of \citet{andrievsky95}.

In the present work, we find that the star A present a $\lambda$ Boo pattern, while the stars B and C present abundances
closer to the Sun. In other words, we find different abundances for the stars A and B, in agreement with \citet{andrievsky95}.
This is difficult to explain under the accretion scenario of $\lambda$ Boo stars.
Then, current evidence does not support the presence of two bonafide $\lambda$ Boo components in this system.

{\bf{HD 38545}}:
\citet{sturenburg93} estimated solar abundances for C (-0.1$\pm$0.2 dex) and almost solar values for other metals
such as Fe (-0.2$\pm$0.2 dex).
However, it was analyzed as a single object and then considered not reliable by \citet{heiter02}.
Then, this object was mentioned as a possible visual binary with a small separation \citep[$<$0.2", ][]{heiter02} and 
as a possible SB system \citep{paunzen02}.
More recently, \citet{prugniel11} reported a low metallicity for this object ([Fe/H]=-0.48 dex) considered also as single.
By inspecting IUE UV spectra, \citet{murphy15} suggest that it is a normal object rather than a $\lambda$ Boo star
("non-member" of the class), and caution that its high $v\sin i$ ($\sim$191 km/s) may also have had some role in early identifications as $\lambda$ Boo.
We note that this star is not included in the list of SB $\lambda$ Boo stars of \citet{paunzen12a}.
To our knowledge, there is no spectral classification nor abundance determination for the secondary.

{\bf{HD 64491}}:
\citet{kamp01} identified this object as a SB system, a previously undetected binary, showing high and low $v\sin i$ components.
They estimated abundances for the star with higher $v\sin i$ ($\sim$ 170 km/s) by directly fitting the composite spectra,
obtaining [N/H]=-0.30 dex, [S/H]=-0.09 dex and [Ca/H]=-0.96 dex (using NLTE for C and S).
Then, \citet{iliev01} reported that the orbital period of this SB system is between 230 and 760 days, and suggest that
a new abundance analysis should be performed taking into account the binarity of the system.
\citet{fg03} suggest that this object is composed by two slightly metal-poor objects ($\sim$-0.5 dex) rather than
a single object with [M/H]$\sim$-1.5 dex.
\citet{murphy15} classified the primary of the system as "F1 Vs kA3mA3 $\lambda$ Boo".
To our knowledge, there is no spectral classification nor abundance determination for the secondary (the object with lower $v\sin i$).

{\bf{HD 84948}}:
\citet{paunzen98} reported this object as a SB system and found subsolar abundances separately
for the stars A and B ([Fe/H]= -1.2$\pm$0.3 dex and -1.0$\pm$0.2 dex, respectively).
Then, \citet{heiter02} also performed a detailed abundance determination separately for components A and B.
Both works reported that that the two stars are metal-poor, however CNO or S abundances were not reported.
Then, \citet{iliev02} estimated NLTE abundances for C and O: they find subsolar values for C (-0.8$\pm$0.4 dex
for both stars) while for O they found -0.6$\pm$0.3 dex and +0.2$\pm$0.3 for stars A and B.
They also reported a period of 7.41 d for this SB2 system.

\begin{figure*}
\centering
\includegraphics[width=8.0cm]{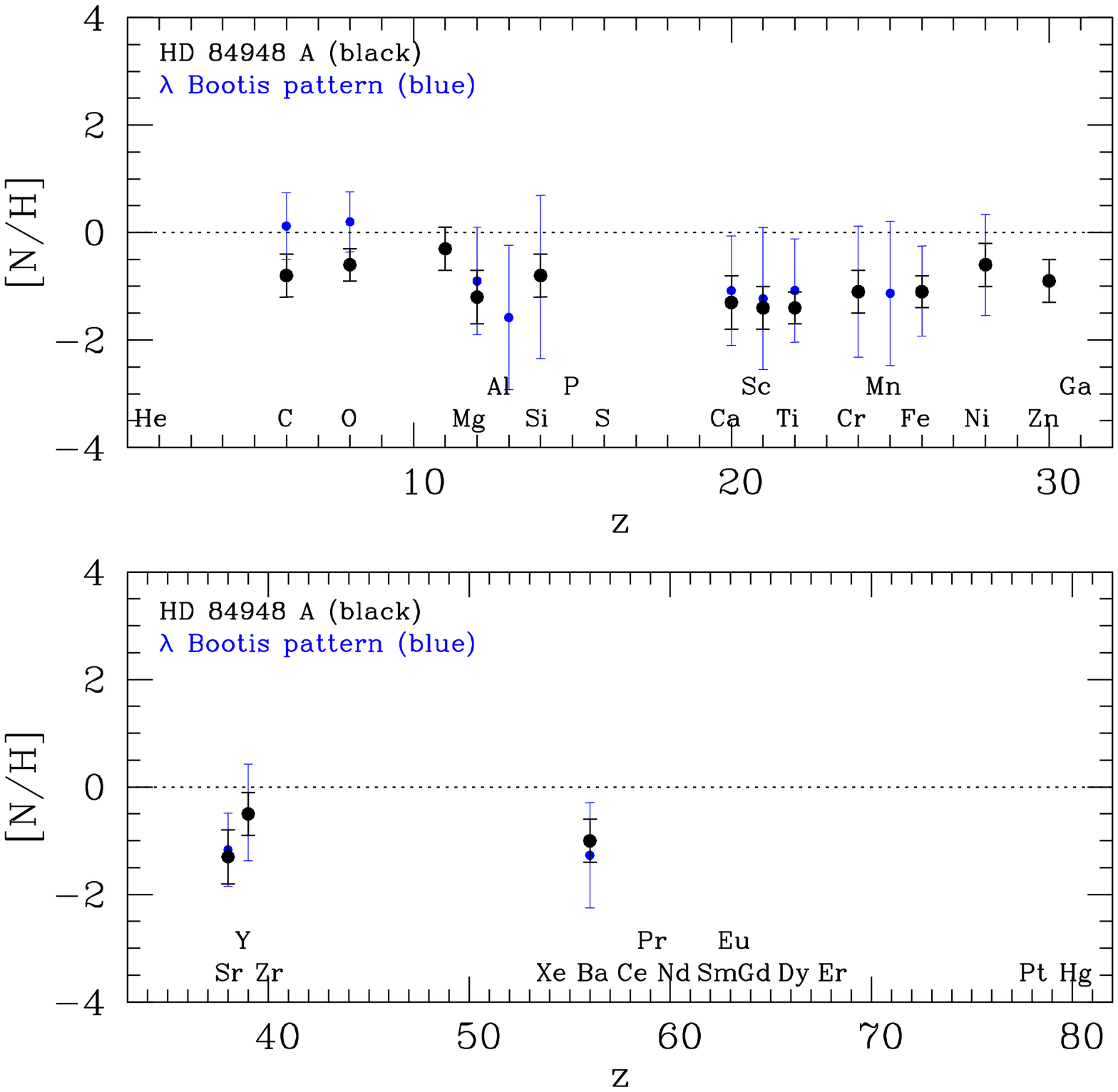}
\includegraphics[width=8.0cm]{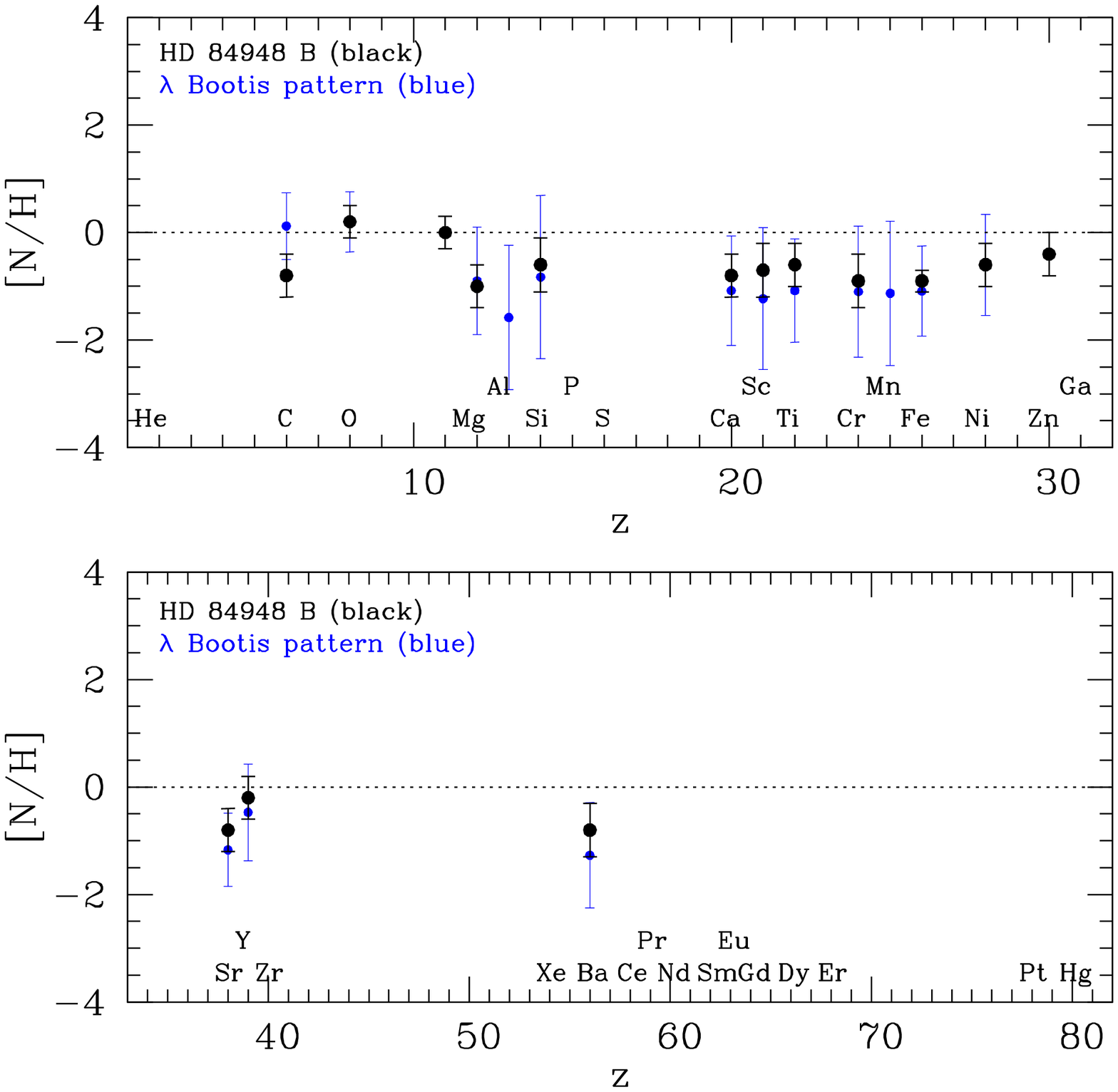}
\caption{Comparison of an average $\lambda$ Boo pattern \citep[blue, ][]{heiter02} with the abundances from literature for the stars 
HD 84948 A and B (left and right panels, black).}
\label{fig.HD84948}
\end{figure*}

We present in the Fig. \ref{fig.HD84948} a comparison of an average $\lambda$ Boo 
pattern\footnote{We excluded from the average stars without CNO values and the stars analyzed here.}
taken from \citet{heiter02}, and literature abundances for the stars A and B.
This plot shows that C abundances seem to be low respect of $\lambda$ Boo stars.
When comparing C with Fe abundances, the group of $\lambda$ Boo stars present [C/Fe]$\sim$1.21$\pm$0.35 dex \citep[excluding stars without CNO values
and the stars analyzed here, ][]{heiter02} with minimum and maximum values of 0.70 and 1.74 dex. 
However, the stars A and B present [C/Fe] values of $\sim$0.4 and $\sim$0.2 
dex\footnote{Using Fe from \citet{heiter02} instead of \citet{paunzen98}, the values are even lower: $\sim$0.3 and $\sim$0.1 dex for stars A and B.},
being low values compared to the average [C/Fe] and even lower than the minimum of 0.70 dex.
These low [C/Fe] values possibly correspond to an extreme or mild-$\lambda$ Boo star rather than to an average $\lambda$ Boo object.

\citet{paunzen01} classified HD 84948 as "kA7hF1mA6 V (LB)", while \citet{murphy15} classified HD 84948 as "F1.5 Vs kA5mA5 $\lambda$ Boo?",
a "probable member" of the $\lambda$ Boo class using a newer spectra.
Given the low values of [C/Fe] for both stars together with the "probable" spectral
classification, we prefer to consider them as candidate $\lambda$ Boo stars (perhaps mild-$\lambda$ Boo stars) rather than bonafide members of the class.
This binary system deserves a verification of the abundance values.

{\bf{HD 111786 (= HR 4881)}}: 
This star is considered as a classic $\lambda$ Boo object by different works \citep[e.g., ][]{murphy15}.
\citet{sturenburg93} derived abundances in agreement with the $\lambda$ Boo class
(for example, [C/H]=-0.2$\pm$0.2 dex and [Fe/H]=-1.5$\pm$0.3 dex).
However, it was analyzed as a single object and then considered not reliable by \citet{heiter02}.
Then, some authors propose an SB nature for this system \citep{faraggiana97,paunzen12b}.
We refer the reader to \citet{murphy15} for a more complete discussion about this object.
The star was classified as "F0 V kA1mA1 $\lambda$ Boo" \citep{murphy15,gray17} and "F0 Vs kA1mA1 $\lambda$ Boo" \citep{murphy20}.
Notably, \citet{faraggiana01} proposed that HD 111786 is in fact a multiple system composed by five members:
one broad-lined star and four narrow-lined stars with similar temperature.
Beyond the multiplicity of this system, to our knowledge there is no spectral classification nor 
abundance determination for the secondary (or any other component) of the system.

{\bf{HD 141851}}: 
\citet{paunzen99} found [C/H] and [O/H] NLTE abundances of -0.81 and -0.21 dex, respectively, showing
$v\sin i$ in excess of 200 km s${^-1}$.
\citet{kamp01} derived LTE abundances of [Ca/H]=-1.30 dex, with typical errors of 0.2 dex.
However, \citet{heiter02} mention that this object was analyzed as a single star and then the abundances are not reliable.
Then, different works claim that this object was misclassified and did not belong to the $\lambda$ Boo class \citep[e.g., ][]{paunzen01}.
\citet{andrievsky02} found [Fe/H] = -0.70, [Si/H] = -0.65 and [Na/H] = +0.60 dex, however they do not decide if this
object is a $\lambda$ Boo star. Then, \citet{murphy15} classified this object as a normal "A2 IVn" star, 
while \citet{gray17} as "A2 IV-Vn", i.e. non-member of the $\lambda$ Boo class.
To our knowledge, there is no spectral classification nor abundance determination for the secondary.

{\bf{HD 148628/638}}: 
The primary of this visual pair (HD 148638) was analyzed by \citet{kamp01} obtaining solar values of N and S
together with subsolar Ca (-1.20 dex). However, \citet{murphy15,murphy20} and \citet{gray17} classified this object as
"A2 IV-n (4481-wk)" and "A2 IVn" rather than a member of the $\lambda$ Boo class. 
To our knowledge, there is no spectral classification nor abundance study for the companion (HD 148628).

{\bf{HD 171948}}: 
Together with HD 84948, \citet{paunzen98} identified this object as the first SB systems with $\lambda$ Boo components.
They reported very low abundances for Mg, Ti, Cr and Fe separately for the components A and B.
Then, \citet{heiter02} derived LTE abundances for this system, estimating the same values within the errors for both stars.
For C they obtained an upper limit ([C/H]<-0.5 dex), while O is considered for the same authors as deficient
([O/H]=-0.6$\pm$0.4 dex) although high compared to heavy elements ([Fe/H]=-1.6$\pm$0.4 dex).
Then, \citet{iliev02} reported NLTE abundances for C and O in this system, estimating the same values for both stars within the errors 
([C/H]=-1.2$\sim$0.4 dex and [O/H]=+0.2$\sim$0.3). They also derived a period of 21.9 days for the SB system.

\begin{figure*}
\centering
\includegraphics[width=8.0cm]{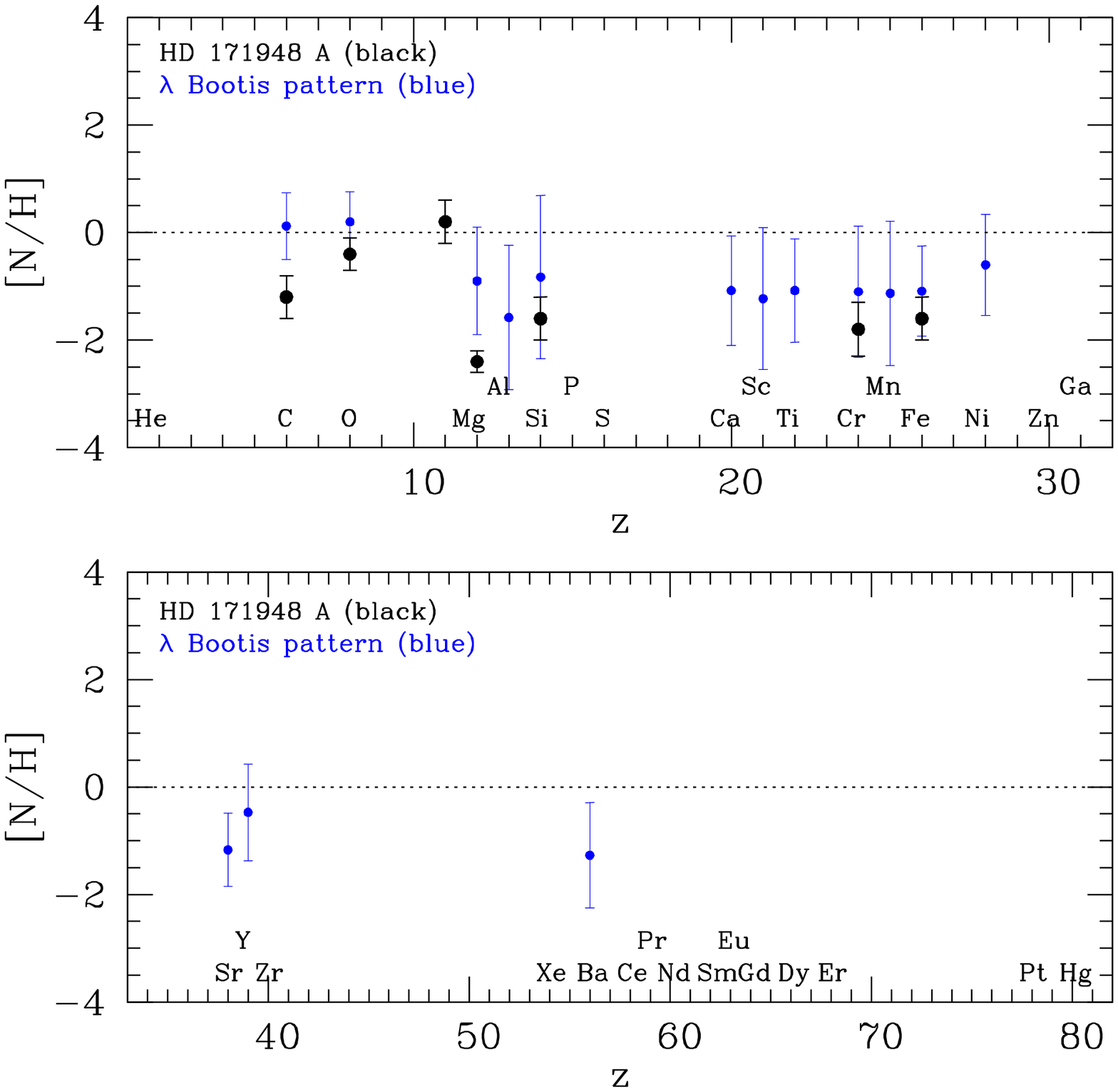}
\includegraphics[width=8.0cm]{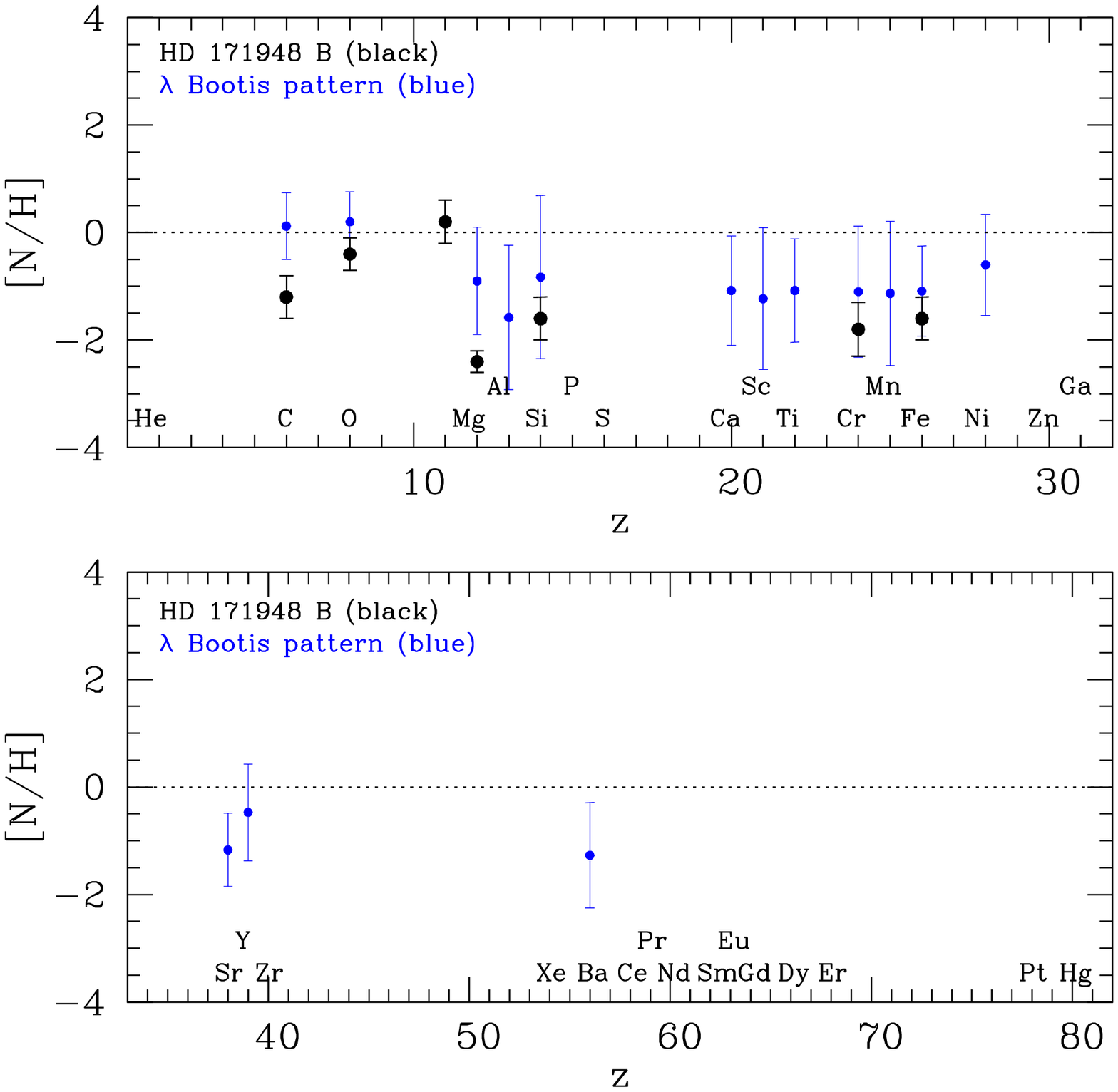}
\caption{Comparison of an average $\lambda$ Boo pattern \citep[blue, ][]{heiter02} with the abundances from literature for the stars 
HD 171948 A and B (left and right panels, black).}
\label{fig.HD171948}%
\end{figure*}

We present in the Fig. \ref{fig.HD171948} a comparison of an average $\lambda$ Boo pattern \citep{heiter02} with the literature
abundances of stars A and B, showing that C values seem to be low respect of $\lambda$ Boo objects.
Comparing C and Fe abundances, both stars A and B present [C/Fe] values of $\sim$0.4 dex (taking NLTE C values from Iliev et al. 2002
and Fe from Heiter et al. 2002), being lower than the average [C/Fe] of $\lambda$ Boo stars
\citep[$\sim$1.21$\pm$0.35 dex excluding stars without CNO values and the stars analyzed here, ][]{heiter02}
and lower than the minimum of 0.70 dex \citep{heiter02}.
We consider that these low [C/Fe] values possibly correspond to an extreme or mild-$\lambda$ Boo star rather than to an average $\lambda$ Boo object.

\citet{murphy15} classified the primary of this binary system as "A3 Va- kB8.5 $\lambda$ Boo", however, there is no spectral classification
listed for the secondary (see their Table 1).
Given the low values of [C/Fe] for both stars and the lack of a spectral classification for the secondary,
we prefer to consider them as candidate $\lambda$ Boo stars (perhaps mild-$\lambda$ Boo stars) rather than bonafide members of the class.
This binary system deserves a verification of the abundance values.

{\bf{HD 174005}}: 
This object was mentioned as a possible SB system with a maximum separation of $\sim$38 arcsec \citep{paunzen00,solano01,paunzen12a}. 
Both \citet{gray01} and \citet{murphy15} classified this object as "A7 V kA2 mA2 $\lambda$ Boo".
To our knowledge, there is no abundance determination for the components of this system, nor
spectral classification for the secondary. This system would deserve a further analysis.

{\bf{HD 193256/281}}: 
The star HD 193281 resulted with near solar C (-0.2$\pm$0.2 dex) and subsolar Fe (-1.0$\pm$0.2 dex),
however also with near solar values of Mg, Ti, Cr and Sr in the study of \citet{sturenburg93}.
Then, \citet{paunzen99} estimated a NLTE oxygen abundance of -0.61 dex.
\citet{kamp01} found solar values in HD 193281 for N, O and S, although for C they found -0.61 dex,
similarly to \citet{paunzen99}.
The star HD 193256 resulted with near solar C (0.0$\pm$0.2 dex) and subsolar Fe (-0.7$\pm$0.2 dex), but also
near solar values of Mg and Si (0.0$\pm$0.2 and 0.0$\pm$0.3 dex) in the study of \citet{sturenburg93}.
Then, abundance values for both stars do not seem to agree with the general pattern of $\lambda$ Boo stars.
The spectra of HD 193256 was classified as "A9 Vn kA2mA2 $\lambda$ Boo" \citep{murphy15} and similarly as
"A8 Vn kA3mA3 ($\lambda$ Boo)" \citep{gray17}.
However, the spectra of HD 193281 was classified as "A2IVn" \citep{murphy15} and "A2 IV-V" \citep{gray17}.
Given a spectral classification in conflict with the abundances, \citet{murphy15} 
consider HD 193281 as an "uncertain member" of the $\lambda$ Boo class.

In this work, we find that HD 193256 present subsolar values of Cr, Mn and Fe, however also near solar values for Mg, Si and Ti, 
which is different than $\lambda$ Boo stars.
For HD 193281, we found a chemical pattern compatible with a a slightly metal deficient star.
However, we also caution that HD 193281 is possibly contaminated by a nearby star (see Sect. 3.2).
Then, current evidence does not support the presence of two bonafide $\lambda$ Boo objects in this system.

{\bf{HD 198160/161}}: 
Both stars were studied separately by \citet{sturenburg93} considering them as twins (same T$_\mathrm{eff}$ and log g),
although \citet{gerbaldi03} criticized this assumption based in their different V and B (0.35 and 0.39 mag).
\citet{sturenburg93} derived near solar values for C in both stars (-0.2$\pm$0.3) and subsolar values for
Fe (-0.8$\pm$0.2 dex), however he also obtained solar values for Mg and Si (0.0$\pm$0.1 dex 
and -0.2$\pm$0.2 dex for both stars).
They also estimated suprasolar values for Na (+0.3$\pm$0.2 dex and +0.6$\pm$0.2 dex for both stars).
Then, \citet{paunzen99} estimated near solar NLTE values for C and O.
\citet{murphy15} classified the spectra of both stars as "A2 Vann $\lambda$ Boo" and "A3 V" (see their Table 1), respectively,
while \citet{gray17} classified the spectra of HD 198160 as "A3 IV(n)".

In this work, we find a general deficiency of metals around 0.7-0.8 dex for both stars.
However, we also found subsolar values for C and O, possibly low compared to other $\lambda$ Boo stars.
When comparing C with Fe abundances, we found that 
the stars HD 198160 and HD 198161 present [C/Fe] values of $\sim$0.54 and $\sim$0.48 dex,
being low compared to the average [C/Fe] of $\lambda$ Boo stars ($\sim$1.21$\pm$0.35 dex) and even lower than the minimum of 0.70 dex (see Sect. 4.3).
Then, we consider that these low [C/Fe] values possibly correspond to mild-$\lambda$ Boo stars,
rather than to an average $\lambda$ Boo object.
In our opinion, current evidence does not support the presence of two bonafide $\lambda$ Boo objects in the system.

{\bf{HD 210111}}: 
\citet{sturenburg93} analyzed this object as a single star, obtaining solar abundances for C (0.1$\sim$0.1 dex),
a subsolar value for Fe (-1.1$\sim$0.2 dex), but also obtaining suprasolar and solar values for Sr and Ba
(+0.45$\sim$0.2 and 0.05$\sim$0.2 dex).
\citet{solano01} obtained subsolar values for Mg, Cr, Sc and Fe (between -0.8 dex and -1.3 dex),
while \citet{kamp01} derived subsolar and near solar abundances for C and O (-0.45 dex and -0.20 dex,
with typical errors of 0.2 dex).
\citet{paunzen99} estimated NLTE values for C and O of -0.45 dex and -0.20 dex.
We suppose that the data presented in these abundance works correspond to the primary of the system,
where its binary nature were not reported.
This object was classified as "kA2hA7mA2 Vas $\lambda$ Boo" with peculiar hydrogen lines by \citet{gray88},
and then as "A9 V kA2mA2 $\lambda$ Boo" by \citet{gray17}.
A classification spectra for HD 210111 was presented by \citet{paunzen12b}, who suggest a SB2 
nature for the system. They fitted the observed data using a composite spectrum with two equal components
having [M/H]=-1.0 dex. For a more detailed abundance analysis, the authors suggest
additional spectra for a large separation of the two components. In particular, 
for the secondary there is no detailed abundance determination (including for the volatile species) nor spectral classification.

\section{Chemical abundances}

We present in this section the chemical abundances derived in this work and their errors.
The total error e$_{tot}$ was derived as the quadratic sum of the line-to-line dispersion e$_{1}$
(estimated as $\sigma/\sqrt{n}$ , where $\sigma$ is the standard deviation),
and the error in the abundances (e$_{2}$, e$_{3}$ and e$_{4}$) when varying T$_{\rm eff}$, $\log g$
and v$_\mathrm{micro}$ by their corresponding uncertainties\footnote{We adopt a minimum of 0.01 dex for the errors e$_{2}$, e$_{3}$ and e$_{4}$.}.
For chemical species with only one line, we adopt as $\sigma$ the standard deviation of iron lines.
Abundance tables show the average abundance and the total error e$_{tot}$, together with
the errors e$_{1}$ to e$_{4}$.

\begin{table}
\centering
\caption{Chemical abundances for HD 15164.}
\begin{tabular}{lrcccc}
\hline
\hline
Specie     & [X/H] $\pm$ e$_{tot}$ & e$_{1}$ & e$_{2}$ & e$_{3}$ & e$_{4}$ \\
\hline
Li I    & 1.32 $\pm$ 0.17 & 0.07 & 0.15 & 0.01 & 0.01 \\ 
C I     & -0.30 $\pm$ 0.05 & 0.02 & 0.04 & 0.02 & 0.01 \\ 
N I     & 0.08 $\pm$ 0.10 & 0.07 & 0.05 & 0.01 & 0.04 \\ 
O I     & 0.12 $\pm$ 0.35 & 0.07 & 0.30 & 0.04 & 0.16 \\ 
Mg I    & -0.12 $\pm$ 0.22 & 0.11 & 0.13 & 0.03 & 0.13 \\ 
Mg II   & 0.08 $\pm$ 0.17 & 0.07 & 0.06 & 0.02 & 0.14 \\ 
Al I    & -0.74 $\pm$ 0.29 & 0.02 & 0.08 & 0.02 & 0.28 \\ 
Si II   & -0.30 $\pm$ 0.14 & 0.04 & 0.06 & 0.02 & 0.12 \\ 
Ca II   & -0.26 $\pm$ 0.17 & 0.07 & 0.15 & 0.01 & 0.02 \\ 
Sc II   & -0.30 $\pm$ 0.27 & 0.17 & 0.07 & 0.03 & 0.19 \\ 
Ti II   & -0.20 $\pm$ 0.16 & 0.02 & 0.08 & 0.02 & 0.14 \\ 
Cr II   & -0.35 $\pm$ 0.08 & 0.02 & 0.03 & 0.02 & 0.07 \\ 
Mn I    & -0.38 $\pm$ 0.16 & 0.04 & 0.14 & 0.01 & 0.06 \\ 
Fe I    & -0.36 $\pm$ 0.15 & 0.01 & 0.05 & 0.01 & 0.14 \\ 
Fe II   & -0.37 $\pm$ 0.11 & 0.01 & 0.03 & 0.01 & 0.11 \\ 
Ni II   & -0.50 $\pm$ 0.10 & 0.07 & 0.06 & 0.02 & 0.02 \\ 
Zn I    & -0.53 $\pm$ 0.12 & 0.02 & 0.12 & 0.01 & 0.01 \\ 
Sr II   & 0.37 $\pm$ 0.32 & 0.02 & 0.16 & 0.01 & 0.27 \\ 
Y II    & -0.26 $\pm$ 0.12 & 0.03 & 0.11 & 0.02 & 0.04 \\ 
Zr II   & -0.06 $\pm$ 0.11 & 0.07 & 0.08 & 0.02 & 0.02 \\ 
Ba II   & 0.10 $\pm$ 0.29 & 0.10 & 0.16 & 0.01 & 0.22 \\ 
\hline
\end{tabular}
\label{tab.abunds.HD15164}
\end{table}

\begin{table}
\centering
\caption{Chemical abundances for HD 15165.}
\begin{tabular}{lrcccc}
\hline
\hline
Specie     & [X/H] $\pm$ e$_{tot}$ & e$_{1}$ & e$_{2}$ & e$_{3}$ & e$_{4}$ \\
\hline
C I     & -0.06 $\pm$ 0.07 & 0.02 & 0.04 & 0.05 & 0.02 \\ 
O I     & 0.52 $\pm$ 0.12 & 0.02 & 0.02 & 0.02 & 0.11 \\ 
Mg I    & -1.06 $\pm$ 0.25 & 0.21 & 0.08 & 0.06 & 0.09 \\ 
Mg II   & -1.00 $\pm$ 0.24 & 0.22 & 0.05 & 0.06 & 0.06 \\ 
Al I    & -1.49 $\pm$ 0.28 & 0.10 & 0.17 & 0.09 & 0.18 \\ 
Ca II   & -1.03 $\pm$ 0.28 & 0.26 & 0.09 & 0.01 & 0.04 \\ 
Sc II   & -1.40 $\pm$ 0.28 & 0.22 & 0.11 & 0.06 & 0.12 \\ 
Ti II   & -0.97 $\pm$ 0.16 & 0.06 & 0.04 & 0.06 & 0.13 \\ 
Cr II   & -1.12 $\pm$ 0.08 & 0.02 & 0.07 & 0.03 & 0.01 \\ 
Fe I    & -1.24 $\pm$ 0.16 & 0.06 & 0.09 & 0.01 & 0.12 \\ 
Fe II   & -1.14 $\pm$ 0.07 & 0.04 & 0.04 & 0.03 & 0.04 \\ 
Sr II   & -0.34 $\pm$ 0.34 & 0.07 & 0.13 & 0.01 & 0.31 \\ 
Ba II   & -0.54 $\pm$ 0.26 & 0.15 & 0.08 & 0.03 & 0.19 \\ 
\hline
\end{tabular}
\label{tab.abunds.HD15165}
\end{table}

\begin{table}
\centering
\caption{Chemical abundances for HD 15165C.}
\begin{tabular}{lrcccc}
\hline
\hline
Specie     & [X/H] $\pm$ e$_{tot}$ & e$_{1}$ & e$_{2}$ & e$_{3}$ & e$_{4}$ \\
\hline
Mg I    & -0.25 $\pm$ 0.11 & 0.10 & 0.01 & 0.01 & 0.01 \\ 
Al I    & -0.08 $\pm$ 0.03 & 0.02 & 0.01 & 0.01 & 0.01 \\ 
Si I    & 0.09 $\pm$ 0.10 & 0.08 & 0.06 & 0.01 & 0.01 \\ 
Ca I    & 0.15 $\pm$ 0.07 & 0.04 & 0.05 & 0.01 & 0.01 \\ 
Sc II   & -0.11 $\pm$ 0.06 & 0.06 & 0.01 & 0.01 & 0.01 \\ 
Ti I    & -0.03 $\pm$ 0.06 & 0.03 & 0.05 & 0.01 & 0.01 \\ 
Ti II   & -0.04 $\pm$ 0.05 & 0.05 & 0.01 & 0.01 & 0.01 \\ 
V I     & 0.01 $\pm$ 0.07 & 0.04 & 0.06 & 0.01 & 0.01 \\ 
Cr I    & -0.02 $\pm$ 0.06 & 0.04 & 0.04 & 0.01 & 0.01 \\ 
Cr II   & -0.02 $\pm$ 0.08 & 0.08 & 0.01 & 0.01 & 0.01 \\ 
Mn I    & 0.29 $\pm$ 0.09 & 0.09 & 0.02 & 0.01 & 0.01 \\ 
Fe I    & 0.04 $\pm$ 0.02 & 0.01 & 0.01 & 0.01 & 0.01 \\ 
Fe II   & -0.01 $\pm$ 0.05 & 0.04 & 0.02 & 0.01 & 0.01 \\ 
Co I    & -0.13 $\pm$ 0.05 & 0.04 & 0.01 & 0.01 & 0.01 \\ 
Cu I    & -0.21 $\pm$ 0.18 & 0.18 & 0.01 & 0.01 & 0.01 \\ 
Zn I    & -0.15 $\pm$ 0.24 & 0.24 & 0.01 & 0.01 & 0.01 \\ 
Sr II   & -0.18 $\pm$ 0.13 & 0.13 & 0.01 & 0.01 & 0.01 \\ 
Y II    & 0.04 $\pm$ 0.21 & 0.21 & 0.01 & 0.01 & 0.03 \\ 
Zr II   & 0.24 $\pm$ 0.13 & 0.13 & 0.01 & 0.02 & 0.01 \\ 
Ba II   & 0.53 $\pm$ 0.17 & 0.17 & 0.01 & 0.01 & 0.01 \\ 
Nd II   & 0.12 $\pm$ 0.08 & 0.08 & 0.01 & 0.01 & 0.01 \\ 
\hline
\end{tabular}
\label{tab.abunds.HD15165C}
\end{table}

\begin{table}
\centering
\caption{Chemical abundances for HD 193256.}
\begin{tabular}{lrcccc}
\hline
\hline
Specie     & [X/H] $\pm$ e$_{tot}$ & e$_{1}$ & e$_{2}$ & e$_{3}$ & e$_{4}$ \\
\hline
C I     & -0.05 $\pm$ 0.22 & 0.21 & 0.04 & 0.02 & 0.05 \\ 
O I     & 0.74 $\pm$ 0.16 & 0.15 & 0.04 & 0.05 & 0.02 \\ 
Mg I    & 0.34 $\pm$ 0.25 & 0.21 & 0.05 & 0.04 & 0.12 \\ 
Mg II   & 0.02 $\pm$ 0.24 & 0.21 & 0.07 & 0.04 & 0.08 \\ 
Si II   & 0.08 $\pm$ 0.18 & 0.07 & 0.16 & 0.04 & 0.02 \\ 
Ca II   & -0.47 $\pm$ 0.23 & 0.21 & 0.06 & 0.05 & 0.01 \\ 
Sc II   & -0.60 $\pm$ 0.31 & 0.21 & 0.08 & 0.03 & 0.21 \\ 
Ti II   & -0.18 $\pm$ 0.25 & 0.07 & 0.03 & 0.08 & 0.23 \\ 
Cr II   & -0.61 $\pm$ 0.09 & 0.02 & 0.02 & 0.07 & 0.06 \\ 
Mn I    & -0.53 $\pm$ 0.13 & 0.04 & 0.10 & 0.01 & 0.07 \\ 
Fe I    & -0.92 $\pm$ 0.15 & 0.07 & 0.06 & 0.02 & 0.12 \\ 
Fe II   & -0.69 $\pm$ 0.10 & 0.04 & 0.04 & 0.05 & 0.08 \\ 
Sr II   & -0.61 $\pm$ 0.49 & 0.27 & 0.10 & 0.10 & 0.38 \\ 
\hline
\end{tabular}
\label{tab.abunds.HD193256}
\end{table}

\begin{table}
\centering
\caption{Chemical abundances for HD 193281.}
\begin{tabular}{lrcccc}
\hline
\hline
Specie     & [X/H] $\pm$ e$_{tot}$ & e$_{1}$ & e$_{2}$ & e$_{3}$ & e$_{4}$ \\
\hline
C I     & -0.35 $\pm$ 0.10 & 0.07 & 0.07 & 0.01 & 0.01 \\ 
O I     & -0.30 $\pm$ 0.06 & 0.03 & 0.04 & 0.02 & 0.01 \\ 
Mg I    & -0.16 $\pm$ 0.29 & 0.20 & 0.10 & 0.06 & 0.18 \\ 
Mg II   & -0.54 $\pm$ 0.21 & 0.18 & 0.02 & 0.01 & 0.11 \\ 
Al I    & -0.65 $\pm$ 0.25 & 0.18 & 0.08 & 0.02 & 0.15 \\ 
Si II   & -0.84 $\pm$ 0.11 & 0.07 & 0.08 & 0.05 & 0.01 \\ 
Ca II   & -0.27 $\pm$ 0.21 & 0.18 & 0.11 & 0.01 & 0.01 \\ 
Sc II   & -0.23 $\pm$ 0.32 & 0.18 & 0.10 & 0.01 & 0.25 \\ 
Ti II   & -0.24 $\pm$ 0.15 & 0.05 & 0.05 & 0.04 & 0.13 \\ 
Cr II   & -0.53 $\pm$ 0.04 & 0.02 & 0.02 & 0.02 & 0.01 \\ 
Fe I    & -0.36 $\pm$ 0.13 & 0.05 & 0.09 & 0.02 & 0.07 \\ 
Fe II   & -0.48 $\pm$ 0.13 & 0.03 & 0.07 & 0.01 & 0.10 \\ 
Sr II   & -0.04 $\pm$ 0.47 & 0.01 & 0.16 & 0.01 & 0.44 \\ 
Y II    & -0.09 $\pm$ 0.16 & 0.13 & 0.09 & 0.04 & 0.01 \\ 
Zr II   & -0.02 $\pm$ 0.19 & 0.18 & 0.06 & 0.02 & 0.01 \\ 
Ba II   & 0.20 $\pm$ 0.17 & 0.09 & 0.14 & 0.01 & 0.03 \\ 
\hline
\end{tabular}
\label{tab.abunds.HD193281}
\end{table}

\begin{table}
\centering
\caption{Chemical abundances for HD 198160.}
\begin{tabular}{lrcccc}
\hline
\hline
Specie     & [X/H] $\pm$ e$_{tot}$ & e$_{1}$ & e$_{2}$ & e$_{3}$ & e$_{4}$ \\
\hline
C I     & -0.29 $\pm$ 0.08 & 0.07 & 0.01 & 0.04 & 0.03 \\ 
O I     & -0.43 $\pm$ 0.28 & 0.15 & 0.02 & 0.02 & 0.23 \\ 
Mg I    & -0.91 $\pm$ 0.18 & 0.08 & 0.03 & 0.01 & 0.15 \\ 
Mg II   & -0.50 $\pm$ 0.20 & 0.15 & 0.06 & 0.05 & 0.10 \\ 
Al I    & -1.25 $\pm$ 0.21 & 0.15 & 0.05 & 0.03 & 0.13 \\ 
Si II   & -0.86 $\pm$ 0.23 & 0.15 & 0.09 & 0.05 & 0.13 \\ 
Ca II   & -0.65 $\pm$ 0.16 & 0.15 & 0.05 & 0.03 & 0.01 \\ 
Sc II   & -0.85 $\pm$ 0.18 & 0.15 & 0.03 & 0.04 & 0.09 \\ 
Ti II   & -0.73 $\pm$ 0.18 & 0.02 & 0.02 & 0.06 & 0.17 \\ 
Cr II   & -0.68 $\pm$ 0.08 & 0.07 & 0.01 & 0.04 & 0.01 \\ 
Mn I    & -1.06 $\pm$ 0.11 & 0.01 & 0.11 & 0.01 & 0.02 \\ 
Fe I    & -0.83 $\pm$ 0.10 & 0.04 & 0.07 & 0.01 & 0.06 \\ 
Fe II   & -0.83 $\pm$ 0.10 & 0.03 & 0.04 & 0.04 & 0.08 \\ 
Sr II   & -1.29 $\pm$ 0.30 & 0.18 & 0.08 & 0.07 & 0.22 \\ 
Ba II   & -0.47 $\pm$ 0.17 & 0.15 & 0.08 & 0.01 & 0.02 \\ 
\hline
\end{tabular}
\label{tab.abunds.HD198160}
\end{table}

\begin{table}
\centering
\caption{Chemical abundances for HD 198161.}
\begin{tabular}{lrcccc}
\hline
\hline
Specie     & [X/H] $\pm$ e$_{tot}$ & e$_{1}$ & e$_{2}$ & e$_{3}$ & e$_{4}$ \\
\hline
C I     & -0.32 $\pm$ 0.06 & 0.04 & 0.02 & 0.04 & 0.02 \\ 
O I     & -0.21 $\pm$ 0.28 & 0.15 & 0.02 & 0.02 & 0.23 \\ 
Mg I    & -0.87 $\pm$ 0.18 & 0.07 & 0.03 & 0.01 & 0.17 \\ 
Mg II   & -0.55 $\pm$ 0.20 & 0.15 & 0.06 & 0.05 & 0.10 \\ 
Al I    & -1.01 $\pm$ 0.25 & 0.15 & 0.06 & 0.04 & 0.18 \\ 
Si II   & -0.38 $\pm$ 0.21 & 0.15 & 0.08 & 0.05 & 0.10 \\ 
Ca II   & -0.67 $\pm$ 0.16 & 0.15 & 0.05 & 0.03 & 0.01 \\ 
Sc II   & -0.85 $\pm$ 0.18 & 0.15 & 0.03 & 0.04 & 0.09 \\ 
Ti II   & -0.83 $\pm$ 0.17 & 0.05 & 0.03 & 0.06 & 0.14 \\ 
Cr II   & -0.68 $\pm$ 0.07 & 0.06 & 0.01 & 0.04 & 0.01 \\ 
Mn I    & -0.97 $\pm$ 0.14 & 0.08 & 0.11 & 0.01 & 0.03 \\ 
Fe I    & -0.80 $\pm$ 0.17 & 0.03 & 0.07 & 0.01 & 0.15 \\ 
Fe II   & -0.81 $\pm$ 0.11 & 0.04 & 0.04 & 0.04 & 0.08 \\ 
Sr II   & -1.41 $\pm$ 0.22 & 0.04 & 0.08 & 0.07 & 0.19 \\ 
Ba II   & -0.38 $\pm$ 0.18 & 0.15 & 0.07 & 0.01 & 0.06 \\ 
\hline
\end{tabular}
\label{tab.abunds.HD198161}
\end{table}

\end{appendix}

\end{document}